\documentclass[11pt]{article}
\usepackage[utf8]{inputenc} 
\usepackage[english]{babel}
\usepackage[T1]{fontenc}
\usepackage[bbgreekl]{mathbbol}
\usepackage[margin=2cm,vmargin=2cm]{geometry}
\usepackage{caption}
\usepackage{cite}
\usepackage{tikz-cd}
\usepackage{amsmath,amsfonts,amssymb,amsthm}
\usepackage{authblk}
\usepackage{csquotes}
\usepackage{url}
\usepackage{fullpage,setspace}
\usepackage{cancel,verbatim,slashed,todonotes}
\usepackage{xcolor,mdframed,graphicx,color}
\usepackage{dsfont}
\usepackage{mathtools}
\usepackage{multirow,stmaryrd}
\usepackage{array}
\usepackage{makecell}
\usepackage{mathrsfs}
\DeclareSymbolFontAlphabet{\mathbbvar}{bbold}
\DeclareSymbolFontAlphabet{\mathbb}{AMSb}
\newcommand{\M}{\mathcal{M}}

\newcommand{\Coo}{\mathcal{C}^{\infty}}

\newcommand{\Diff}{\mathrm{Diff}}

\usepackage{slashed}

\newcommand{\di}{\mathrm{d}}

\usepackage{upgreek}

\DeclareRobustCommand\longtwoheadrightarrow
    {\relbar\joinrel\twoheadrightarrow}
\DeclareRobustCommand\longhookrightarrow
     {\lhook\joinrel\longrightarrow}
\newcommand{\xtwoheadrightarrow}[2][]{%
  \mathrel{\ooalign{$\xrightarrow[#1\mkern4mu]{#2\mkern4mu}$\cr%
  \hidewidth$\rightarrow\mkern4mu$}}
}
\theoremstyle{definition}
\newtheorem{theorem}{Lemma}[section]
\theoremstyle{definition}

\theoremstyle{definition}
\newtheorem{definition}[theorem]{Definition}
\theoremstyle{definition}
\newtheorem{remark}[theorem]{Remark}
\theoremstyle{definition}
\newtheorem{example}[theorem]{Example}
\theoremstyle{definition}
\newtheorem{notation}[theorem]{Notation}
\newmdtheoremenv[linecolor=black, linewidth=1pt,leftmargin=-10,
rightmargin=-10,backgroundcolor=white,%
innertopmargin=0pt,%
splittopskip=\topskip,skipbelow=\baselineskip,%
skipabove=\baselineskip,ntheorem]{theorembox}%
[theorem]{Lemma}
\newmdtheoremenv[linecolor=black, linewidth=1pt,leftmargin=-10,
rightmargin=-10,backgroundcolor=white,%
innertopmargin=0pt,%
splittopskip=\topskip,skipbelow=\baselineskip,%
skipabove=\baselineskip,ntheorem]{theoremboxT}%
[theorem]{Theorem}
\numberwithin{equation}{subsection}
\tikzcdset{%
    triple line/.code={\tikzset{%
        double equal sign distance, 
        double=\pgfkeysvalueof{/tikz/commutative diagrams/background color}}},
    quadruple line/.code={\tikzset{%
        double equal sign distance, 
        double=\pgfkeysvalueof{/tikz/commutative diagrams/background color}}},
    Rrightarrow/.code={\tikzcdset{triple line}\pgfsetarrows{tikzcd implies cap-tikzcd implies}},
    RRightarrow/.code={\tikzcdset{quadruple line}\pgfsetarrows{tikzcd implies cap-tikzcd implies}}
}

\makeatletter
\providecommand{\leftsquigarrow}{%
  \mathrel{\mathpalette\reflect@squig\relax}%
}
\newcommand{\reflect@squig}[2]{%
  \reflectbox{$\m@th#1\rightsquigarrow$}%
}
\makeatother
\usepackage{hyperref}
\title{\vspace{-1cm}
\begin{flushleft}
\small{Prepared for submission to Complex Manifolds,\\ special issue: Generalized Geometry} \vspace{-1.15cm}
\end{flushleft}
\begin{flushright}
\normalsize{QMUL-PH-21-11}
\end{flushright}
\vspace{2cm}
\bigskip
\bf
Towards an extended/higher correspondence\\
{\large Generalised geometry, bundle gerbes and global Double Field Theory}
\vspace{1cm}}
\author{\sc Luigi Alfonsi}
\affil{\em\normalsize Centre for Research in String Theory,\\\em School of Physics and Astronomy,\\\em Queen Mary University of London,\\\em 327 Mile End Road, London E1 4NS, UK\\\vspace{4mm}\tt \href{mailto:l.alfonsi@qmul.ac.uk}{l.alfonsi@qmul.ac.uk}}
\date{}

\begin{document}
\maketitle

\vspace{0.5cm}
\abstract{
\noindent In this short paper, we will review the proposal of a correspondence between the doubled geometry of Double Field Theory and the higher geometry of bundle gerbes.
Double Field Theory is T-duality covariant formulation of the supergravity limit of String Theory, which generalises Kaluza-Klein theory by unifying metric and Kalb-Ramond field on a doubled-dimensional space.
In light of the proposed correspondence, this doubled geometry is interpreted as an atlas description of the higher geometry of bundle gerbes. In this sense, Double Field Theory can be interpreted as a field theory living on the total space of the bundle gerbe, just like Kaluza-Klein theory is set on the total space of a principal bundle.
This correspondence provides a higher geometric interpretation for para-Hermitian geometry which opens the door to its generalisation to Exceptional Field Theory. \vspace{0.2cm}

\noindent This review is based on, but not limited to, my talk at the workshop \textit{Generalized Geometry and Applications} at Universit\"{a}t Hamburg on 3rd of March 2020.

\vspace{0.3cm}

\vspace{0.3cm}
\noindent \textbf{MSC classes}: \tt{53C08}, \tt{53D18}, \tt{83E30}
}

\newpage
\tableofcontents
\section{Introduction}\label{s1}

One of the most characteristic and fascinating features of String Theory, when compared to the usual field theories, is the appearance of T-duality: an additional, hidden symmetry of the theory. Double Field Theory is a T-duality covariant formulation of the supergravity limit of String Theory which makes this symmetry manifest. Double Field Theory was proposed in \cite{HulZwi09} and seminal work includes \cite{Siegel1, Siegel2}. See \cite{BerTho14, BerBla20} for reviews. As enlightened by \cite{Ber19,BerBla20}, Double Field Theory can be interpreted as a generalisation of Kaluza-Klein theory, which geometrically unifies the metric with the Kalb-Ramond field, instead of a gauge field.

\paragraph{The higher geometry of T-duality.}
The Kalb-Ramond field is, geometrically, the connection of a bundle gerbe $\mathscr{G}\twoheadrightarrow M$, a categorification of a $U(1)$-bundle, which was introduced by \cite{Murray, Murray2} and reformulated in terms of \v{C}ech cohomology by \cite{Hit99}. In \cite{Principal1}, bundle gerbes are formalised as principal $\infty$-bundles in the context of higher geometry.
Given a good cover $\{U_\alpha\}$ of the base manifold $M$, the connection of a bundle gerbe is given by local $2$-forms $B_{(\alpha)}\in\Omega^2(U_\alpha)$, local $1$-forms $\Lambda_{(\alpha\beta)}\in\Omega^1(U_\alpha\cap U_\beta)$ and local scalars $G_{(\alpha\beta\gamma)}\in\Coo(U_\alpha\cap U_\beta \cap U_\gamma)$, which are patched on overlaps of patches by
\begin{equation}\label{eq:introgerby}
    \begin{aligned}
    H \,&=\, \mathrm{d}B_{(\alpha)} \\
    B_{(\beta)} - B_{(\alpha)} \,&=\, \mathrm{d}\Lambda_{(\alpha\beta)} \\
    \Lambda_{(\alpha\beta)}+\Lambda_{(\beta\gamma)}+\Lambda_{(\gamma\alpha)} \,&=\, \mathrm{d}G_{(\alpha\beta\gamma)} \\
    G_{(\alpha\beta\gamma)}-G_{(\beta\gamma\delta)}+G_{(\gamma\delta\alpha)}-G_{(\delta\alpha\beta)} \,&\in\, 2\pi\mathbb{Z}.
    \end{aligned}
\end{equation}

\noindent Since the Kalb-Ramond field is the connection of a bundle gerbe, T-duality has been naturally formulated in the context of higher geometry. Topological T-duality \cite{Bou03,Bou03x,Bou03xx,Bou04} is based on the topological properties of bundle gerbes and T-duality has been formulated as a particular isomorphism of bundle gerbes in \cite{Bunke:2005um, BunNik13, FSS16x, FSS17x,FSS18,FSS18x,NikWal18}. We will now briefly introduce such a formulation, by unravelling \cite[Definition 2.8]{Bunke:2005um}. \vspace{0.2cm}

\noindent Let us consider two $T^n$-bundle spacetimes $M\xrightarrow{\pi}M_0$ and $\widetilde{M}\xrightarrow{\widetilde{\pi}}M_0$ over a common base manifold $M_0$, with first Chern classes $c_1(M)\in H^2(M_0,\mathbb{Z})^n$ and $\widetilde{c}_1(\widetilde{M})\in H^2(M_0,\mathbb{Z})^n$. Then, consider the couple of bundle gerbes $\mathscr{G}\xrightarrow{\Pi}M$ and $\widetilde{\mathscr{G}}\xrightarrow{\widetilde{\Pi}}\widetilde{M}$, encoding two Kalb-Ramond fields respectively on $M$ and $\widetilde{M}$, with Dixmier-Douady classes of the form
\begin{equation}
    [H] \,=\, \left[ \sum_{i=1}^n h_i\otimes \widetilde{c}_1(\widetilde{M})^i\right]\in H^3(M,\mathbb{Z}), \qquad [\widetilde{H}] \,=\, \left[ \sum_{i=1}^n \widetilde{h}^i\otimes c_1(M)_i\right]\in H^3(\widetilde{M},\mathbb{Z}),
\end{equation}
where $h_i$ and $\widetilde{h}^i$ are respectively the generators of the cohomology of the fibres $T^n$ and $\widetilde{T}^n$.
Then, the bundle gerbes $\mathscr{G}$ and $\widetilde{\mathscr{G}}$ are geometric T-dual if there exists an isomorphism  
\begin{equation}
    \begin{tikzcd}[row sep={11ex,between origins}, column sep={12ex,between origins}]
    & \mathscr{G}\times_{M_0} \widetilde{M}\arrow[rr, "\cong"', "\mathrm{T\text{-}duality}"]\arrow[dr, "\Pi"']\arrow[dl, "\widetilde{\pi}"] & & M\times_{M_0}\widetilde{\mathscr{G}}\arrow[dr, "\pi"']\arrow[dl, "\widetilde{\Pi}"] \\
    \mathscr{G}\arrow[dr, "\Pi"'] & & M\times_{M_0}\widetilde{M}\arrow[dr, "\pi"']\arrow[dl, "\widetilde{\pi}"] & & [-2.5em]\widetilde{\mathscr{G}}\arrow[dl, "\widetilde{\Pi}"] \\
    & M\arrow[dr, "\pi"'] & & \widetilde{M}\arrow[dl, "\widetilde{\pi}"] & \\
    & & M_0 & &
    \end{tikzcd}
\end{equation}
such that the following condition, known as \textit{Poincar\'e condition}, is satisfied: for any given point $x\in M_0$, we must have
\begin{equation}
    \big[\mathrm{T\text{-}duality}|_x\big] \,=\, \left[ \sum_{i=1}^n h_i \smile \widetilde{h}^i \right] \,\in\, \frac{H^2(T^n \times \widetilde{T}^n,\mathbb{Z})}{\!\mathrm{im}(\pi^\ast|_x)\oplus\mathrm{im}(\widetilde{\pi}^\ast|_x)},
\end{equation}
where we used the fact that an isomorphism of bundle gerbes is equivalently a $U(1)$-bundle on its base manifold. \vspace{0.2cm}

\noindent This diagram is closely related to the diagram in \cite{CavGua11}, which formalises T-duality in the context of generalised geometry:
\begin{equation}
    \begin{tikzcd}[row sep={14ex,between origins}, column sep={15ex,between origins}]
     & TK\oplus T^\ast K\arrow[dr, "\pi_\ast"']\arrow[dl, "\widetilde{\pi}_\ast"] \\
     TM\oplus T^\ast M\arrow[dr, "/T^n"'] & & [-2.6em]T\widetilde{M}\oplus T^\ast \widetilde{M}\arrow[dl, "/\widetilde{T}^n"] \\
     & TM_0\oplus T^\ast M_0 \oplus (M_0\times \mathbb{R}^{2n}) & 
    \end{tikzcd}
\end{equation}
where we called $K:=M\times_{M_0}\widetilde{M}$. This is because we can interpret the Courant algebroid as a geometric object which embodies the infinitesimal symmetries of a bundle gerbe \cite{Gua11}.
In this sense T-duality is a geometric property of bundle gerbes.

\subsection{Introduction to local Double Field Theory}

Here, we will give a brief introduction to the formalism of local Double Field Theory.

\paragraph{Doubled patch.}
Let us consider an open simply connected $2d$-dimensional patch $\mathcal{U}$. We can introduce coordinates $(x^\mu,\widetilde{x}_\mu):\mathcal{U}\rightarrow \mathbb{R}^{2d}$, which we will call collectively $x^M:=(x^\mu,\widetilde{x}_\mu)$. Now, we want to equip the vector space $\mathbb{R}^{2d}$ with the fundamental representation of the continuous T-duality group $O(d,d)$. Since the action of $O(d,d)$-matrices on $\mathbb{R}^{2d}$ preserves the matrix $\eta_{MN}:=\left(\begin{smallmatrix}0&1\\1&0\end{smallmatrix}\right)$, we can define a metric $\eta=\eta_{MN}\di x^M \otimes \di x^N \in \odot^2T^\ast\mathcal{U}$ with signature $(d,d)$. Here, $\odot$ is the symmetric product defined by $A_1\odot \cdots \odot A_n := \frac{1}{n!}\sum_{\sigma\in S_n}\!A_1\otimes \cdots \otimes A_n$, where $S_n$ is the symmetric group on $n$ symbols.

\paragraph{Gauge algebra.}
We want now to define a generalised Lie derivative which preserves the $\eta$-tensor, i.e. such that $\mathfrak{L}_X\eta=0$ for any vector field $X\in\mathfrak{X}(\mathcal{U})$. Thus, for any couple of vector fields $X,Y\in\mathfrak{X}(\mathcal{U})$ we can define
\begin{equation}
    \big(\mathfrak{L}_XY\big)^M \;:=\;  X^N\partial_N Y^M - \mathbb{P}^{ML}_{\quad\; NP} \partial_LX^NY^P,
\end{equation}
where we defined the tensor
\begin{equation}
    \mathbb{P}^{ML}_{\quad\; NP} \,:=\, \delta^M_P\delta^L_N  - \eta^{ML}\eta_{NP},
\end{equation}
which projects the $GL(2d)$-valued function $\partial_LX^N$ into an $\mathfrak{o}(d,d)$-valued one. The generalised Lie derivative is also known as D-bracket $\llbracket X,Y \rrbracket_{\mathrm{D}} := \mathfrak{L}_XY$.
The {C-bracket} is defined as the anti-symmetrisation of the D-bracket, i.e.
\begin{equation}
    \llbracket X,Y \rrbracket_{\mathrm{C}} \;:=\; \frac{1}{2}\big(\llbracket X,Y \rrbracket_{\mathrm{D}} - \llbracket Y,X \rrbracket_{\mathrm{D}}\big).
\end{equation}

\noindent Now, if we want to construct an algebra of generalised Lie derivatives, we immediately find out that it cannot be close, i.e. we generally have
\begin{equation}
    \big[\mathfrak{L}_X, \,\mathfrak{L}_Y\big] \;\neq\;\mathfrak{L}_{\llbracket X,Y \rrbracket_{\mathrm{C}}}
\end{equation}
Thus, to assure the closure, we need to impose extra conditions. The {weak} and the  {strong constraint} (also known collectively as  {section condition}) are respectively the conditions
\begin{equation}
    \eta^{MN}\partial_M\partial_N\phi_i =0, \qquad  \eta^{MN}\partial_M\phi_1\partial_N\phi_2 =0
\end{equation}
for any couple of fields or parameters $\phi_1,\phi_2$. The immediate solution to the section condition is obtained by considering only fields and parameters $\phi$ which satisfy the condition $\widetilde{\partial}^\mu\phi=0$. Therefore, upon application of the strong constraint, all the fields and parameters will depend on the $d$-dimensional quotient manifold $U := \mathcal{U}/\!\sim \,\,\hookrightarrow \mathcal{U}$, where $\sim$ is the relation identifying points with the same physical coordinates $(x^\mu,\widetilde{x}_\mu)\sim(x^\mu,\widetilde{x}_\mu')$. In particular, vector fields $X\in\mathfrak{X}(\mathcal{U})$ satisfying the strong constraint can be identified with sections of the generalised tangent bundle $TU\oplus T^\ast U$ of generalised geometry. Moreover the C-bracket, when restricted to strong constrained vectors, reduces to the Courant bracket of generalised geometry, i.e. we have
\begin{equation}
    \llbracket -,- \rrbracket_{\mathrm{C}}\,\Big|_{\widetilde{\partial}^\mu=0} \;=\; [-,-]_{\mathrm{Cou}}
\end{equation}
In this sense, the geometry underlying Double Field Theory, when strong constrained, locally reduces to generalised geometry. 

\paragraph{Generalised metric.} 
We can define the {generalised metric} $\mathcal{G}=\mathcal{G}_{MN}\di x^M\otimes \di x^N$ by requiring that it is symmetric and it satisfies the property $\mathcal{G}_{ML}\eta^{LP}\mathcal{G}_{PN}=\eta_{MN}$. Thus, the matrix $\mathcal{G}_{MN}$ can be parametrised as
\begin{equation}
    \mathcal{G}_{MN} \;=\; \begin{pmatrix}g_{\mu\nu}- B_{\mu\lambda}g^{\lambda\rho}B_{\rho\beta} & B_{\mu\lambda}g^{\lambda\nu} \\-g^{\mu\lambda}B_{\lambda\nu} & g^{\mu\nu} \end{pmatrix}.
\end{equation}
where $g_{\mu\nu}$ and $B_{\mu\nu}$ are respectively a symmetric and an anti-symmetric matrix. 
Finally, we must impose the strong constraint on $\mathcal{G}_{MN}$, so that its components are allowed to depend only on the $x^\mu$ coordinates, and not on the $\widetilde{x}_\mu$ ones. 
Now, $g:=g_{\mu\nu}\di x^\mu\otimes \di x^\nu$ is a symmetric tensor and $B:=\frac{1}{2}B_{\mu\nu}\di x^\mu\wedge \di x^\nu$ is an anti-symmetric tensor on the $d$-dimensional quotient manifold $U$. These can be respectively interpreted as a metric and a Kalb-Ramond field on the $d$-dimensional patch $U$. 
If we consider a strong constrained vector $V:=v+\widetilde{v}\in\mathfrak{X}(U)\oplus \Omega^1(U)$. The infinitesimal gauge transformation given by generalised Lie derivative $\delta \mathcal{G}_{MN}=\mathfrak{L}_V\mathcal{G}_{MN}$ is equivalent to the following gauge transformations:
\begin{equation}
    \delta g = \mathcal{L}_v g, \qquad \delta B = \mathcal{L}_v B + \di\widetilde{v}
\end{equation}
where $\mathcal{L}_v$ is the ordinary Lie derivative. This, then reproduces the gauge transformations of metric and Kalb-Ramond field. Therefore, the infinitesimal generalised diffeomorphisms of the $2d$-dimensional patch $\mathcal{U}$ unify the infinitesimal diffeomorphisms of the $d$-dimensional patch $U\hookrightarrow \mathcal{U}$ with the infinitesimal gauge transformations of the Kalb-Ramond field, in analogy with Kaluza-Klein theory.

\paragraph{The globalisation problem.}
However, the Kalb-Ramond field $B$ is geometrically the connection of a bundle gerbe and hence it is globalised by the patching conditions \eqref{eq:introgerby}. Thus, it is not obvious how the local patches $\big(\mathcal{U},\,\sim,\,\eta,\,\mathcal{G}\big)$, which we introduced here, can be consistently glued together? This is the substance of the globalisation problem of the doubled geometry underlying Double Field Theory. Seminal work in this direction was done by \cite{BCM14}. The purpose of this paper is to try to answer this question. 
\vspace{0.2cm}

\noindent We know how to globalise the local geometry of Double Field Theory for particular classes of examples, where the gerby nature of the Kalb-Ramond field is not manifest. In particular, global Double Field Theory on group manifolds \cite{Hul09, DFTWZW15, DFTWZW15x, Hass18} is well-defined. Also, doubled torus bundles \cite{Hull06}, which are globally affine $T^{2n}$-bundles on an undoubled base manifold \cite{BelHulMin07}, are well-defined. However, there is no conclusive answer on how to globalise this geometry in the most general case. Moreover, it has been argued in \cite{HohSam13} that the doubled torus bundles should be recoverable by imposing a certain compactified topology to a general doubled space, whose geometry, however, remains an open problem. A problem which becomes even more obscure in the case of the geometry underlying Exceptional Field Theory.

\subsection{Para-Hermitian geometry for Double Field Theory}\label{para}

The first proposal of formalisation of the geometry underlying Double Field Theory as a para-K\"{a}hler manifold was developed by \cite{Vais12} and then generalised to a para-Hermitian manifold by \cite{Vai13}. The para-Hermitian program was further developed by \cite{Svo17, Svo18, MarSza18, Svo19, MarSza19, Shiozawa:2019jul, Hassler:2019wvn, BPV20, Ba20, Ikeda:2020lxz, Svo20}.

\paragraph{Para-complex geometry.}
An almost para-complex manifold $(\M,J)$ is a $2d$-dimensional smooth manifold $\M$ which is equipped with a $(1,1)$-tensor field $J\in\mathrm{End}(T\M)$, called almost para-complex structure, such that $J^2=\mathrm{id}_{T\mathcal{M}}$ and that the $\pm 1$-eigenbundles $L_\pm\subset T\mathcal{M}$ of $J$ have both $\mathrm{rank}(L_\pm)=d$. An almost para-complex structure is, then, equivalently given by a splitting of the form
\begin{equation}
    T\mathcal{M}\;=\;L_+\oplus L_-
\end{equation}
Therefore, the structure group of the tangent bundle $T\mathcal{M}$ of the almost para-complex manifold is reduced to $GL(d,\mathbb{R})\times GL(d,\mathbb{R})\subset GL(2d,\mathbb{R})$. The para-complex structure also canonically defines the following projectors to its eigenbundles:
\begin{equation}
    \Pi_\pm \,:=\, \frac{1}{2}(1\pm J): \,T\M \,\longtwoheadrightarrow\, L_\pm.
\end{equation}
An almost para-complex structure $J$ is said to be, respectively, $\pm$-integrable if $L_\pm$ is closed under Lie bracket, i.e. if it satisfies the property
\begin{equation}
    \big[\Gamma(\mathcal{M},L_\pm),\,\Gamma(\mathcal{M},L_\pm)\big]_{\mathrm{Lie}} \,\subseteq\, \Gamma(\mathcal{M},L_\pm).
\end{equation}
The $\pm$-integrability of $J$ implies the existence a foliation $\mathcal{F}_\pm$ of the manifold $\mathcal{M}$ such that $L_\pm = T \mathcal{F}_\pm$. An almost para-complex manifold $(\mathcal{M},J)$ is a para-complex manifold if and only if $J$ is both $+$-integrable and $-$-integrable at the same time.
\vspace{0.25cm}

\paragraph{Para-Hermitian geometry.}
An almost para-Hermitian manifold $(\mathcal{M},J,\eta)$ is an almost para-complex manifold $(\mathcal{M},J)$ equipped with a metric $\eta \in \bigodot^2T^\ast\mathcal{M}$ of split signature $(d,d)$ which is compatible with the almost para-complex structure as it follows:
\begin{equation}
    \eta(J-,J-) \,=\, - \eta(-,-).
\end{equation}
A para-Hermitian structure $(J,\eta)$ canonically defines an almost symplectic structure $\omega\in\Omega^2(\mathcal{M})$, called fundamental $2$-form, by $\omega(-,-) := \eta(J-,-)$. An almost para-Hermitian manifold can be equivalently expressed as $(\mathcal{M},J,\omega)$, since the para-Hermitian metric can be uniquely determined by $\eta(-,-) = \omega(J-,-)$. Notice that the subbundles $L_\pm$ are both maximal isotropic subbundles respect to $\eta$ and Lagrangian subbundles respect to $\omega$. 

\paragraph{Recovering generalised geometry.}
The para-Hermitian metric immediately induces an isomorphism $\eta^\sharp:L_{\pm}\xrightarrow{\;\cong\;}L_{\mp}^\ast$. In the case of a $+$-integrable para-Hermitian manifold, this implies the existence of an isomorphism
\begin{equation}
    T\mathcal{M}\;\cong\;T\mathcal{F}_+\oplus T^\ast\mathcal{F}_+
\end{equation}
given by $X\mapsto \Pi_+(X) + \eta^\sharp(\Pi_-(X))$, for any vector $X\in T\M$.
As shown by \cite{Svo17,MarSza18}, it is possible to define a bracket structure $\llbracket-,-\rrbracket_{\mathrm{D}}: \mathfrak{X}(\mathcal{M})\times\mathfrak{X}(\mathcal{M})\rightarrow\mathfrak{X}(\mathcal{M})$ which is compatible with the para-Hermitian metric, so that $(T\M,\llbracket-,-\rrbracket_{\mathrm{D}},\eta)$ is a metric algebroid, and which makes a generalised version of the Nijenhuis tensor of $J$ vanish \cite[p.$\,$13]{MarSza18}.
If we consider any couple of sections $X+\xi,Y+\zeta\in\Gamma(\M,T\mathcal{F}_+\oplus T^\ast\mathcal{F}_+)$, the bracket can be rewritten as
\begin{equation}
    \llbracket X+\xi, Y+\zeta\rrbracket_{\mathrm{D}} \;=\!\! \underbrace{\big([X,Y] + \mathcal{L}_X\zeta - \iota_Y\di\xi  \big)}_{\text{Dorfman bracket on }T\mathcal{F}_+\oplus T^\ast\mathcal{F}_+}\!\!\! + \;\big( [\xi,\zeta]^\ast + \mathcal{L}_\xi^\ast Y - \iota_\zeta\di^\ast X \big)
\end{equation}
where $[-.-]^\ast$, $\mathcal{L}^\ast_{(-)}$ and $\di^\ast$ are operators induced by the Lie bracket of $T\M$. Therefore, if we restrict ourselves to couples of strongly foliated vectors, i.e. $X+\xi,Y+\zeta\in\mathfrak{X}(\mathcal{F}_+)\oplus\Omega^1(\mathcal{F}_+)$, we recover the usual Dorfman bracket
\begin{equation}
    \llbracket X+\xi, Y+\zeta\rrbracket_{\mathrm{D}} \;=\; [X,Y] + \mathcal{L}_X\zeta - \iota_Y\di\xi,
\end{equation}
i.e. we recover generalised geometry.
\vspace{0.25cm}

\noindent An almost para-Hermitian manifold $(\mathcal{M},J,\eta)$ is, in particular, a para-K\"{a}hler manifold if the fundamental $2$-form is symplectic, i.e. $\di\omega=0$. In the general case, the closed $3$-form $\mathcal{K}\in\Omega^3_{\mathrm{cl}}(\mathcal{M})$ defined by $\mathcal{K}:=\di\omega$, which embodies the obstruction of $\omega$ from being symplectic, is interpreted as the generalised fluxes of Double Field Theory.

\paragraph{Born geometry.}
A Born geometry is the datum of an almost para-Hermitian manifold $(\mathcal{M},J,\omega)$ equipped with a Riemannian metric $\mathcal{G}\in \bigodot^2T^\ast\mathcal{M}$ which is compatible with both the metric $\eta$ and the fundamental $2$-form $\omega$ as it follows:
\begin{equation}
    \eta^{-1}\mathcal{G} \,=\, \mathcal{G}^{-1}\eta \quad\;\text{and}\;\quad \omega^{-1}\mathcal{G} \,=\, -\mathcal{G}^{-1}\omega.
\end{equation}
Such a Riemannian metric can be identified with the generalised metric of Double Field Theory.

\paragraph{Generalised T-dualities.}
As explained by \cite{MarSza19}, the generalised diffeomorphisms of Double Field Theory can now be identified with diffeomorphisms of $\mathcal{M}$ which preserve the para-Hermitian metric $\eta$, i.e isometries $\mathrm{Iso}(\mathcal{M},\eta)$. Thus, at any point $p\in\mathcal{M}$, the push-forward $f_\ast|_p:T_p\mathcal{M}\rightarrow T_{f(p)}\mathcal{M}$ of a generalised diffeomorphism $f\in\mathrm{Iso}(\mathcal{M},\eta)$ is given by an $O(d,d)\subset GL(2d)$ transformation. The Jacobian of such a diffeomorphism can then be seen as an $O(d,d)$-valued function, which we will call $f_\ast\in\Coo(\mathcal{M},O(d,d))$, by a slight abuse of notation. 
\vspace{0.2cm}

\noindent This group of symmetries can be further extended to the group of general bundle automorphisms of $T\mathcal{M}$ preserving the para-Hermitian metric $\eta$.
Thus, a generalised diffeomorphism induces a morphism of Born geometries
\begin{equation}
    (\mathcal{M},\,J,\,\omega,\,\mathcal{G})\;\longmapsto\;(\mathcal{M},\, f^\ast J,\,f^\ast\omega,\, f^\ast\mathcal{G}),
\end{equation}
which is an isometry of the para-Hermitian metric, i.e. such that it preserves $\eta = f^\ast\eta$. 
\vspace{0.2cm}

\noindent Particularly interesting is the case of the $b$-shift, which can be interpreted as a bundle morphism $e^{b}:T\mathcal{M}\rightarrow T\mathcal{M}$ covering the identity $\mathrm{id}_{\mathcal{M}}$ of the base manifold. Given $b\in\wedge^2L^\ast_+$, we can define this bundle morphism by 
\begin{equation}
    e^b(X) \;=\; X_+ + b^\sharp(X_+) + X_-,
\end{equation}
for any vector $X=X_++X_-$ with components $X_\pm\in L_\pm$.
This transforms the para-complex structure by $J\mapsto J+b^\sharp$, which also implies $\omega\mapsto \omega+b$. Therefore, a $b$-shift maps the splitting $T\mathcal{M} = L_+\oplus L_-$ to a new one $T\mathcal{M} =L_+^\prime\oplus L_-$, preserving the eigenbundle $L_-$, but not $L_+$. Therefore, it does not preserve $+$-integrability.

\paragraph{The patching puzzle of para-Hermitian geometry.} However, as shown in \cite{Alf20}, if we are interested in recovering a general conventional geometric background, given by a general spacetime manifold $M$ equipped with a general bundle gerbe connection $(B_{(\alpha)}, \Lambda_{(\alpha\beta)}, G_{(\alpha\beta\gamma)})$, we encounter a conceptual problem.\vspace{0.2cm}

\noindent If we want to consider a conventional bosonic supergravity background, there must exists a foliation $\mathcal{F}_-$ of $\mathcal{M}$ such that $L_- = T\mathcal{F}_-$ and the leaf space $M:=\mathcal{M}/\mathcal{F}_-$ of this foliation must be a smooth manifold. Thus, the foliation $\mathcal{F}_-$ is simple and the canonical quotient map $\pi:\mathcal{M}\twoheadrightarrow M= \mathcal{M}/\mathcal{F}_-$ is a surjective submersion, making $\mathcal{M}$ a fibered manifold. Let now $(\widetilde{x}_{\mu},x^\mu)$ be local coordinates adapted to the foliation $\mathcal{F}_-$, i.e. fibered, on any patch $\mathcal{U}_\alpha$. Then, the fundamental $2$-form $\omega\in\Omega^2(\M)$ must have the form \cite[p.$\,$40]{MarSza19}
\begin{equation}
    \omega \;=\; \di \widetilde{x}_{(\alpha)\mu}\wedge\di x^\mu_{(\alpha)} - \pi^\ast B_{(\alpha)}.    
\end{equation}
Since, this satisfies $\pi^\ast H = \di (\pi^\ast B_{(\alpha)}  ) = -\di\omega$, where $H\in\Omega^3_\mathrm{cl}(M)$ is the curvature of the Kalb-Ramond field, we would expect to be possible for the local $2$-forms $B_{(\alpha)}$ to be patched as a general connection of a bundle gerbe. In other words, by defining the patches of the leaf space by $U_{\alpha}:=\pi\!\left(\mathcal{U}_{\alpha}\right)$, we would expect the following general patching conditions
\begin{equation}\label{eq:introgerbe}
    \begin{aligned}
    B_{(\beta)} - B_{(\alpha)} &= \di \Lambda_{(\alpha\beta)} &&\text{ on } U_{\alpha}\cap U_{\beta}, \\
    \Lambda_{(\alpha\beta)} + \Lambda_{(\beta\gamma)} + \Lambda_{(\gamma\alpha)} &=\di G_{(\alpha\beta\gamma)}  &&\text{ on }U_{\alpha}\cap U_{\beta}\cap U_{\gamma},\\
    G_{(\alpha\beta\gamma)} + G_{(\beta\alpha\delta)} + G_{(\gamma\beta\delta)} + G_{(\delta\alpha\gamma)} &\in 2\pi\mathbb{Z} &&\text{ on }U_{\alpha}\cap U_{\beta}\cap U_{\gamma}\cap U_{\delta}
    \end{aligned}
\end{equation}
to be allowed.
However, as shown by \cite{Alf20}, the transition functions of $\mathcal{M}$ on two-fold overlaps of patches $\mathcal{U}_{\alpha}\cap\mathcal{U}_{\beta}$ force, in general, the bundle gerbe to be trivial. Therefore, if we want to embed a conventional supergravity background into an almost para-Hermitian manifold, we have some troubles.\vspace{0.2cm}

\noindent Thus, we have two main questions to answer about para-Hermitian geometry. Firstly, why does it work so well? We are currently unable to provide a well-defined generalisation of para-Hermitian geometry to Exceptional Field Theories, but if we could be able to derive para-Hermitian geometry from a more fundamental geometric principle, perhaps this would give us the key to find such a generalisation. Secondly, how can we modify its globalisation such that recovering a conventional supergravity background becomes possible?
The formalism proposed in \cite{Alf19, Alf20} tries to answer both questions. To explain how, we first need to introduce more elements of higher geometry.

\section{Bundle gerbes}\label{s2}

In this section we will briefly introduce some fundamental notions in higher geometry, with particular focus on bundle gerbes. For an introductory self-contained review, see \cite{Bunk:2021quu}.\vspace{0.2cm}

\noindent Let $\mathbf{Diff}$ be the ordinary category of smooth manifolds and $\mathbf{\infty Grpd}$ the $(\infty,1)$-category of $\infty$-groupoids. Roughly, a smooth stack $\mathscr{X}$ is defined as an $\infty$-functor
\begin{equation}
    \mathscr{X}:\mathbf{Diff}^{\mathrm{op}}\longrightarrow\mathbf{\infty Grpd}
\end{equation}
which satisfies some higher gluing properties, known as descent. For more details, we redirect to \cite{DCCTv2}. This can be thought of as a generalisation of the notion of sheaf which takes value in Lie $\infty$-groupoids.
We will call $\mathbf{H}$ the $(\infty,1)$-category of smooth stacks on manifolds. 

\begin{example}[Manifolds as smooth stacks]
Given any smooth manifold $M\in\mathbf{Diff}$, we can easily construct a sheaf $\Coo(-,M)\in\mathbf{H}$ of smooth functions to $M$, which is in particular a stack. This is nothing but a Yoneda embedding $\mathbf{Diff}\hookrightarrow\mathbf{H}$ of the smooth manifolds into the $(\infty,1)$-category of stacks.
\end{example}

\noindent The abelian bundle gerbe is a categorification of the principal $U(1)$-bundle introduced by \cite{Murray, Murray2}. More recently, in \cite{Principal1}, the bundle gerbe has been reformalised as a special case of principal $\infty$-bundle, where the structure Lie $2$-group is $G=\mathbf{B}U(1)$, i.e. the circle $2$-group.

\begin{definition}[Circle $2$-group]
The circle $2$-group $\mathbf{B}U(1)\in\mathbf{H}$ is defined as the group-stack which sends a smooth manifold $M$ to the groupoid $\mathbf{B}U(1)(M)$ whose objects are $U(1)$-bundles on $M$ and whose morphisms are bundle isomorphisms.
The group-stack structure is given by the following bundle isomorphisms
\begin{equation}\label{eq:stacksu}
    \begin{aligned}
    &P^{-1}\otimes P\,\cong\, M\times U(1), \quad P\otimes P^{-1}\,\cong\, M\times U(1), \\
    &P\otimes(P'\otimes P'')\,\cong\,(P\otimes P')\otimes P''
    \end{aligned}
\end{equation}
where, for any given circle bundle $P$, we called $P^{-1}$ its dual bundle, which is a circle bundle with opposite $1$st Chern class, i.e. such that $\mathrm{c}_1(P^{-1})=-\mathrm{c}_1(P)$.
\end{definition}

\noindent Thus, the tensor product $\otimes$ plays the role of the group multiplication, the trivial bundle $M\times U(1)$ plays the role of the {identity element} and $P^{-1}$ plays the role of the {inverse element} of $P$.

\noindent Let us now give a concrete description of this geometrical object.

\begin{definition}[Bundle gerbe]
Let $M$ be the smooth manifold that we can identify with usual spacetime. A bundle gerbe is defined as a principal $\mathbf{B}U(1)$-bundle $\mathscr{G}\xtwoheadrightarrow{\;\Pi\;} M$ by the following pullback diagram in the $(\infty,1)$-category $\mathbf{H}$ of higher smooth stacks:
\begin{equation}\begin{tikzcd}[row sep={12ex,between origins}, column sep={13ex,between origins}]
\mathscr{G} \arrow[d, "\Pi"', two heads]\arrow[r] &\ast \arrow[d]  \\
M\arrow[r, "f"] &\mathbf{B}^2U(1)
\end{tikzcd}\end{equation}
where the higher stack $\mathbf{B}^2U(1):=\mathbf{B}(\mathbf{B}U(1))$ is the delooping of the group-stack $\mathbf{B}U(1)$ and the map $f:M\rightarrow\mathbf{B}^2U(1)$ is the \v{C}ech cocycle of the bundle gerbe. 
\end{definition}

\noindent In this paper, the two-headed arrow $\twoheadrightarrow$ is used to denote epimorphisms and the hooked arrow $\hookrightarrow$ to denote monomorphisms.

\begin{remark}[Bundle gerbe in local data]
Let $\mathcal{U}:=\{U_\alpha\}$ be any good cover for the base manifold $M$. The \v{C}ech groupoid $\check{C}(\mathcal{U})$ is defined as the $\infty$-groupoid corresponding to the following simplicial object
\begin{equation}\label{eq:cechgroupoidsimplicial}
    \begin{tikzcd}[row sep=scriptsize, column sep=6ex] \cdots\arrow[r, yshift=1.8ex, two heads]\arrow[r, yshift=0.6ex, two heads]\arrow[r, yshift=-0.6ex, two heads]\arrow[r, yshift=-1.8ex, two heads] & \bigsqcup_{\alpha\beta\gamma}U_{\alpha}\cap U_\beta\cap U_\gamma\arrow[r, yshift=1.4ex, two heads]\arrow[r, two heads]\arrow[r, yshift=-1.4ex, two heads]& \bigsqcup_{\alpha\beta}U_{\alpha}\cap U_\beta  \arrow[r, yshift=0.7ex, two heads] \arrow[r, yshift=-0.7ex, two heads] & \; \bigsqcup_{\alpha} U_{\alpha} \arrow[r, two heads] & \check{C}(\mathcal{U}).
    \end{tikzcd}
\end{equation}
Now, by using the natural equivalence between the \v{C}ech groupoid $\check{C}(\mathcal{U})$ and the manifold $M$ in the $(\infty,1)$-category of stacks, we can express the map between $M$ and the moduli stack $\mathbf{B}^2U(1)$ as a functor of the form
\begin{equation}
\begin{tikzcd}[row sep=scriptsize, column sep=8ex] 
    M \;\simeq\; \check{C}(\mathcal{U}) \arrow[r, "f"] & \mathbf{B}^2U(1).
\end{tikzcd}
\end{equation}
By using the definition of the \v{C}ech groupoid, such a map can be presented as a collection of cocycles $\bigsqcup_{\alpha\beta}U_{\alpha}\cap U_\beta\rightarrow \mathbf{B}U(1)$ which are glued by isomorphisms on three-fold overlaps of patches $\bigsqcup_{\alpha\beta\gamma}U_{\alpha}\cap U_\beta\cap U_\gamma$. Since, as we have seen, any map $U\rightarrow\mathbf{B}U(1)$ from an open set $U$ is equivalently a $U(1)$-bundle $P\twoheadrightarrow U$, we obtain the following diagram:
\begin{equation}
    \begin{tikzcd}[row sep={11ex,between origins}, column sep={6ex}]
    \big\{\mu_{(\alpha\beta\gamma)}:P_{\alpha\beta}\otimes P_{\beta\gamma}\xrightarrow{\;\cong\;}P_{\alpha\gamma}\big\} \arrow[d, two heads] & \bigsqcup_{\alpha\beta}P_{\alpha\beta} \arrow[d, two heads] & \\
    \bigsqcup_{\alpha\beta\gamma}U_{\alpha}\cap U_\beta\cap U_\gamma \arrow[r, yshift=1.4ex, two heads]\arrow[r, two heads]\arrow[r, yshift=-1.4ex, two heads] & \bigsqcup_{\alpha\beta}U_{\alpha}\cap U_\beta  \arrow[r, yshift=0.7ex, two heads] \arrow[r, yshift=-0.7ex, two heads] & \; \bigsqcup_{\alpha} U_{\alpha} \arrow[d, two heads] \\
    & & M
    \end{tikzcd}
\end{equation}
More in detail, we have a collection of circle bundles $\{P_{\alpha\beta}\twoheadrightarrow U_\alpha\cap U_\beta\}$ on each overlap of patches $U_\alpha\cap U_\beta\subset M$ such that:
\begin{itemize}
    \item there exists a bundle isomorphism $P_{\alpha\beta}\cong P^{-1}_{\beta\alpha}$ on any two-fold overlap of patches $U_\alpha\cap U_\beta$,
    \item there exists a bundle isomorphism $\mu_{(\alpha\beta\gamma)}:P_{\alpha\beta}\otimes P_{\beta\gamma}\xrightarrow{\;\cong\;} P_{\alpha\gamma}$ on any three-fold overlap of patches $U_\alpha\cap U_\beta\cap U_\gamma$,
    \item this isomorphism satisfies $\mu_{(\alpha\beta\gamma)}\circ\mu_{(\beta\gamma\delta)}^{-1}\circ\mu_{(\gamma\delta\alpha)}^{-1}\circ\mu_{(\delta\alpha\beta)}= 1$ on any four-fold overlaps of patches $U_\alpha\cap U_\beta\cap U_\gamma \cap U_\delta$.
\end{itemize}
We, thus, recovered the Hitchin-Chatterjee formulation \cite{Hit99} of the bundle gerbe $\Pi:\mathscr{G}\longtwoheadrightarrow M$. 
\end{remark}

\begin{remark}[Topological classification of bundle gerbes]
Notice that the trivialisation we introduced defines \v{C}ech cocycle $\big[G_{(\alpha\beta\gamma)}\big]\in H^{2}\big(M,\,\Coo(-,S^1)\big)$, where $\Coo(-,S^1)$ is the sheaf of maps to the circle. It is a well-established result (e.g. see see \cite{Hit99} for details) that there exists an isomorphism $H^{2}\big(M,\,\Coo(-,S^1)\big)\cong H^{3}(M,\mathbb{Z})$, induced by the short exact sequence of sheaves $0\rightarrow \mathbb{Z} \rightarrow \Coo(-,\mathbb{R}) \rightarrow \Coo(-,S^1) \rightarrow 0$. The image of $\big[G_{(\alpha\beta\gamma)}\big]$ along such an isomorphism will be an element of the $3$rd cohomology group of the base manifold $M$, which we will call Dixmier-Douady class $\mathrm{dd}(\mathscr{G})\in H^{3}(M,\mathbb{Z})$ of the bundle gerbe. Thus, bundle gerbes $\mathscr{G}\in\mathbf{H}$ are topologically classified by a Dixmier-Douady class $\mathrm{dd}(\mathscr{G})\in H^{3}(M,\mathbb{Z})$.
\end{remark}

\begin{theorem}[Automorphisms of the bundle gerbe]\label{th:autom}
As seen by \cite{BMS20}, the $2$-group of automorphisms of a bundle gerbe $\mathscr{G}\xtwoheadrightarrow{\;\Pi\;} M$ is
\begin{equation}
    \mathrm{Aut}(\mathscr{G}) \;=\; \Diff(M)\,\ltimes\, \mathbf{H}(M,\mathbf{B}U(1))
\end{equation}
where $\mathbf{H}(M,\mathbf{B}U(1)) = \mathbf{B}U(1)(M)$ is the $2$-group of $U(1)$-bundles on $M$. 
\end{theorem}


\subsection{Bundle gerbes with connective structure}

\noindent Let $\mathbf{B}U(1)_{\mathrm{conn}}\in\mathbf{H}$ be the stack of $U(1)$-bundles with connection.

\begin{definition}[Bundle gerbe with connective structure]
Let $M$ be the smooth manifold that we can identify with usual spacetime. A bundle gerbe with connective structure is defined as a principal $\mathbf{B}U(1)_{\mathrm{conn}}$-bundle $\mathscr{G}\xtwoheadrightarrow{\;\Pi\;} M$ by the following pullback diagram in the $(\infty,1)$-category $\mathbf{H}$ of higher smooth stacks:
\begin{equation}\begin{tikzcd}[row sep={14ex,between origins}, column sep={27ex,between origins}]
\mathscr{G} \arrow[d, "\Pi"', two heads]\arrow[r] &\ast \arrow[d]  \\
M\arrow[r, "{(\Lambda_{(\alpha\beta)},G_{(\alpha\beta\gamma)})}"] &\mathbf{B}(\mathbf{B}U(1)_{\mathrm{conn}})
\end{tikzcd}\end{equation}
where the higher stack $\mathbf{B}(\mathbf{B}U(1)_{\mathrm{conn}})$ is the delooping of the group-stack $\mathbf{B}U(1)_{\mathrm{conn}}$ and the map $(\Lambda_{(\alpha\beta)},G_{(\alpha\beta\gamma)}):M\rightarrow\mathbf{B}(\mathbf{B}U(1)_{\mathrm{conn}})$ is the \v{C}ech cocycle of the bundle gerbe with connective structure. 
\end{definition}

\noindent Such a cocycle is given by a collection of local $1$-forms $\Lambda_{(\alpha\beta)}\in\Omega^1(U_\alpha\cap U_\beta)$ and local scalars $G_{(\alpha\beta\gamma)}\in\Coo(U_\alpha\cap U_\beta \cap U_\gamma)$ such that:
\begin{equation}
    \begin{aligned}
    \Lambda_{(\alpha\beta)}+\Lambda_{(\beta\gamma)}+\Lambda_{(\gamma\alpha)} \,&=\, \mathrm{d}G_{(\alpha\beta\gamma)}, \\
    G_{(\alpha\beta\gamma)}-G_{(\beta\gamma\delta)}+G_{(\gamma\delta\alpha)}-G_{(\delta\alpha\beta)} \,&\in\, 2\pi\mathbb{Z}.
    \end{aligned}
\end{equation}

\begin{theorem}[Automorphisms of the bundle gerbe with connective structure]
We can now refine lemma \ref{th:autom} to a bundle gerbe $\mathscr{G}$ with connective structure.
As explained in \cite{FSS16}, the $\infty$-groupoid of automorphisms of the bundle gerbe $\mathscr{G}$ with connective structure
\begin{equation}
    \mathrm{Aut}(\mathscr{G}) \;=\; \Diff(M)\,\ltimes\, \mathbf{H}(M,\mathbf{B}U(1)_{\mathrm{conn}})
\end{equation}
This is nothing but the higher geometric version of the gauge group $G_{\mathrm{DFT}}\;=\;\Diff(M)\ltimes \Omega^2_{\mathrm{cl}}(M)$ of DFT proposed by \cite{Hull14}. In fact, the natural map $\mathbf{H}(M,\mathbf{B}U(1)_{\mathrm{conn}}) \rightarrow \Omega^2_{\mathrm{cl}}(M)$ is just the curvature map sending a $U(1)$-bundle to its curvature $b\in\Omega^2_{\mathrm{cl}}(M)$.
\end{theorem}

\begin{definition}[Connection of a bundle gerbe]
Let $M$ be a smooth manifold. A bundle gerbe with connection is given by a cocycle
\begin{equation}\begin{tikzcd}[row sep={13ex,between origins}, column sep={30ex,between origins}]
M\arrow[r, "{(B_{(\alpha)},\Lambda_{(\alpha\beta)},G_{(\alpha\beta\gamma)})}"] &\mathbf{B}^2U(1)_{\mathrm{conn}},
\end{tikzcd}\end{equation}
where the stack $\mathbf{B}^2U(1)_{\mathrm{conn}}\in\mathbf{H}$ is defined as follows:
\begin{equation}
    \mathbf{B}^2U(1)_{\mathrm{conn}}:\,U \longmapsto\, \mathbf{H}\big( \mathscr{P}(U),\,\mathbf{B}U(1) \big)
\end{equation}
for any differential manifold $U$ and where $ \mathscr{P}(U)$ is the path $\infty$-groupoid of $U$.
The cocycle  $(B_{(\alpha)},\Lambda_{(\alpha\beta)},G_{(\alpha\beta\gamma)}):M\rightarrow\mathbf{B}(\mathbf{B}U(1)_{\mathrm{conn}})$ is given by a collection of local $2$-forms $B_{(\alpha)}\in\Omega^1(U_\alpha\cap U_\beta)$, local $1$-forms $\Lambda_{(\alpha\beta)}\in\Omega^1(U_\alpha\cap U_\beta)$ and local scalars $G_{(\alpha\beta\gamma)}\in\Coo(U_\alpha\cap U_\beta \cap U_\gamma)$ such that:
\begin{equation}
    \begin{aligned}
    B_{(\beta)} - B_{(\alpha)} \,&=\, \mathrm{d}\Lambda_{(\alpha\beta)}, \\
    \Lambda_{(\alpha\beta)}+\Lambda_{(\beta\gamma)}+\Lambda_{(\gamma\alpha)} \,&=\, \mathrm{d}G_{(\alpha\beta\gamma)}, \\
    G_{(\alpha\beta\gamma)}-G_{(\beta\gamma\delta)}+G_{(\gamma\delta\alpha)}-G_{(\delta\alpha\beta)} \,&\in\, 2\pi\mathbb{Z}.
    \end{aligned}
\end{equation}
\end{definition}

\noindent Notice that there is an obvious couple of forgetful functors:
\begin{equation}
    \mathbf{B}^2U(1)_{\mathrm{conn}} \,\longtwoheadrightarrow\, \mathbf{B}(\mathbf{B}U(1)_{\mathrm{conn}}) \,\longtwoheadrightarrow\, \mathbf{B}^2U(1).
\end{equation}

\section{Global Double Field Theory on bundle gerbes}\label{s4}

In this section we will construct the correspondence between the doubled geometry of Double Field Theory and the higher geometry of bundle gerbes. We will define an atlas of a bundle gerbe and we will show that it can be identified with the doubled space of Double Field Theory. This will have the consequence that Double Field Theory can be globally interpreted as a field theory on the total space of a bundle gerbe, just like ordinary Kaluza-Klein theory lives on the total space of a principal bundle. 

\begin{definition}[$0$-truncation of stacks]
Let $\mathbf{H}_0$ be the ordinary category of sheaves on manifolds. Then, the inclusion $\mathbf{H}_0\hookrightarrow \mathbf{H}$ has a left adjoint $\tau_0:\mathbf{H}\rightarrow \mathbf{H}_0$ which is called $0$-truncation and which sends a higher stack $\mathscr{X}\in\mathbf{H}$ to its restricted sheaf $\tau_0\mathscr{X}\in\mathbf{H}_0$ at the $0$-degree.
\end{definition}

\begin{definition}[Atlas of a smooth stack]
An atlas of a smooth stack $\mathscr{X}\in\mathbf{H}$ is defined by a smooth manifold $\mathcal{A }$ equipped with a morphism of smooth stacks
\begin{equation}
    \Phi:\,\mathcal{A }\; \longrightarrow\; \mathscr{X}
\end{equation}
which is, in particular, an effective epimorphism, i.e. whose $0$-truncation $\tau_0\Phi:\mathcal{A }\longtwoheadrightarrow \tau_0\mathscr{X}$ is an epimorphism of sheaves.
\end{definition}

\noindent This formalizes the idea that to a smooth stack $\mathscr{X}\in\mathbf{H}$ we can associate an atlas which is made up of ordinary manifolds $\mathcal{A }$. This provides a remarkably handy tool to deal with higher geometric objects. See \cite{Hei05} and \cite{topos} for more detail. Moreover, the notion of atlas will be a pivotal in establishing a relation between doubled and higher geometry.

\begin{remark}[Gluing morphisms of stacks]
Given a smooth stack $\mathscr{X}\in\mathbf{H}$ equipped with an atlas $\Phi:\mathcal{A } \longrightarrow \mathscr{X}$, we can write the \v{C}ech nerve of $\Phi$ as the following simplicial object
\begin{equation}
\begin{tikzcd}[column sep=6ex]
\dots \arrow[r, yshift=2ex, two heads]\arrow[r, yshift=0.7ex, two heads]\arrow[r, yshift=-0.7ex, two heads] \arrow[r, yshift=-2ex, two heads] &\mathcal{A } \times_\mathscr{X} \mathcal{A } \times_\mathscr{X} \mathcal{A } \arrow[r, yshift=1.4ex, two heads]\arrow[r, two heads] \arrow[r, yshift=-1.4ex, two heads] &\mathcal{A } \times_\mathscr{X} \mathcal{A } \arrow[r, yshift=0.7ex, two heads] \arrow[r, yshift=-0.7ex, two heads] & \mathcal{A }  \arrow[r, "\Phi"] & \mathscr{X} .
\end{tikzcd}
\end{equation}
For simplicity, let us now consider just a geometric $1$-stack $\mathscr{X}\in\mathbf{H}$. A complicated object such as a morphism of stacks $\underline{\sigma}:\mathscr{X}\longrightarrow\mathscr{S}$, for some $\mathscr{S}\in\mathbf{H}$, can be equivalently expressed on the atlas $\mathcal{A }$ of the stack $\mathscr{X}$. This can be done as the map induced by the atlas
\begin{equation}
\begin{tikzcd}[column sep=7ex, row sep=7ex]
\mathcal{A } \arrow[r, "\Phi"]\arrow[rr, bend right=35, "\sigma"]& \mathscr{X} \arrow[r, "\underline{\sigma}"] & \mathscr{S}
\end{tikzcd}
\end{equation}
together with an isomorphism of the two maps induced by the kernel pair of the atlas
\begin{equation}
\begin{tikzcd}[column sep=7ex, row sep=7ex]
\mathcal{A } \times_\mathscr{X} \mathcal{A } \arrow[bend left=40, "\sigma"]{r}[name=U,below]{}
\arrow[bend right=40, "\sigma'"']{r}[name=D]{} &
\mathscr{S} \arrow[Rightarrow, to path=(U) -- (D)]{}
\end{tikzcd}
\end{equation}
such that it satisfies the cocycle condition on $\mathcal{A } \times_\mathscr{X} \mathcal{A } \times_\mathscr{X} \mathcal{A }$. For more details, see \cite{Hei05}.
\end{remark}

\noindent The idea of gluing morphisms of stacks on an atlas will be useful in this section, when we will have to consider geometric structures on a bundle gerbe.

\subsection{The $\mathfrak{double}/\mathfrak{string}$ correspondence}

The aim of this section will be to prove the existence of a correspondence between doubled and higher geometry in a linearised form. 

\begin{definition}[$\mathfrak{string}$ Lie $2$-algebra]
Let us call $\mathfrak{string}:=\mathbb{R}^{d}\oplus\mathbf{b}\mathfrak{u}(1)$ the $2$-algebra of the abelian Lie $2$-group $\mathbb{R}^{d}\times\mathbf{B}U(1)$. It is well-understood that any $L_\infty$-algebra $\mathfrak{g}$ is equivalently described in terms of its Chevalley-Eilenberg dg-algebra $\mathrm{CE}(g)$. In our particular case this is
\begin{equation}
    \mathrm{CE}\!\left(\mathfrak{string}\right) \;=\; \mathbb{R}[e^a,B]/\langle\di e^a=0,\;\di B = 0\rangle,
\end{equation}
where $\{e^a\}$ with $a=0,\dots,d-1$ are generators in degree $1$ and $B$ is a generator in degree $2$.
\end{definition}

\noindent The Lie $2$-algebra $\mathfrak{string}=\mathbb{R}^{d}\oplus\mathbf{b}\mathfrak{u}(1)\twoheadrightarrow \mathbb{R}^{d}$ can be interpreted as a linearisation of a bundle gerbe, in the sense proposed by \cite{Fiorenza:2013nha}. Thus, such a Lie $2$-algebra can be thought as trivially made up of a flat Minkowski space and a trivial Kalb-Ramond field. 
Now, we want to introduce a notion of atlas for this $2$-algebra.

\begin{remark}[Transgression element]
A transgression element of a cocycle $\mu\in\mathrm{CE}({\mathfrak{g}})$ on a fibration $\widehat{\mathfrak{g}}\xtwoheadrightarrow{\;\Pi\;}\mathfrak{g}$ is defined as an element $B\in\mathrm{CE}(\widehat{\mathfrak{g}})$ on such that $\di B = \Pi^\ast(\mu)$.
Morever, it can be proved (see \cite{FSS18}) that, if the fibration is the higher central extension
\begin{equation}
\begin{tikzcd}[row sep={11.5ex,between origins}, column sep={12ex,between origins}]
    \widehat{\mathfrak{g}} \arrow[d, "\mathrm{hofib}(\mu)"', two heads]\arrow[r] & \ast \arrow[d]  \\
    \mathfrak{g} \arrow[r, "\mu"] & \mathbf{b}^n\mathfrak{u}(1),
\end{tikzcd}
\end{equation}
then such a transgression element is universal. This means exactly that, given a transgression element $\omega\in\mathrm{CE}(\mathfrak{h})$ of $\mu\in\mathrm{CE}({\mathfrak{g}})$ for another fibration $\mathfrak{h}\xtwoheadrightarrow{\;\pi\;}\mathfrak{g}$, then there is a unique morphism of fibrations
\begin{equation}
\begin{tikzcd}[row sep={9.5ex,between origins}, column sep={9ex,between origins}]
    \mathfrak{h} \arrow[dr, "\pi"', two heads]\arrow[rr, "\phi"] & & \widehat{\mathfrak{g}} \arrow[dl, "\mathrm{hofib}(\mu)", two heads]  \\
    & \mathfrak{g} & 
\end{tikzcd}
\end{equation}
such that $\omega = \phi^\ast(B)$.
\end{remark}

\begin{remark}[Atlas of an $L_\infty$-algebra]\label{def:atlasAlg}
By linearising the notion of atlas of a smooth stack \cite{DCCTv2, khavkine2017synthetic}, we obtain that an atlas of an $L_\infty$-algebra $\mathfrak{g}$ can be defined by an ordinary Lie algebra $\mathfrak{atlas}$ equipped with a homomorphism of $L_\infty$-algebras $\phi:\mathfrak{atlas} \longrightarrow \mathfrak{g}$ that is surjective onto the $0$-truncation of $\mathfrak{g}$.
In this dissertation, we will also require that the dual homomorphism $\phi^\ast:\mathrm{CE}(\mathfrak{g})\hookrightarrow\mathrm{CE}(\mathfrak{atlas})$ of dg-algebras is injective.
\end{remark}

\noindent We will also need the following slight specialisation of the notion of atlas of an $L_\infty$-algebra, which will be useful to deal with our physically motivated examples.

\begin{definition}[Lorentz-compatible atlas]\label{def:lorentzcom}
Let $\mathfrak{g}\twoheadrightarrow \mathbb{R}^{d}$ be an $L_\infty$-algebra fibrated on a Minkowski space and equipped with an atlas $\phi:\mathfrak{atlas} \longrightarrow \mathfrak{g}$. We say that the atlas is \textit{Lorentz-compatible} if $\mathfrak{atlas}$ comes equipped with a $SO(1,d-1)$-action such that
\begin{enumerate}
    \item it non-trivially extends the natural $SO(1,d-1)$-action on $\mathbb{R}^d$,
    \item $\phi$ is $SO(1,d-1)$-equivariant,
\end{enumerate}
and if $\mathrm{dim}(\mathfrak{atlas})$ is the minimal dimension for which the above conditions are satisfied.
\end{definition}

\noindent Notice that, in a Lorentz-compatible atlas, the images of the higher generators of $\mathfrak{g}$ through the map $\phi^\ast:\mathrm{CE}(\mathfrak{g})\hookrightarrow\mathrm{CE}(\mathfrak{atlas})$ are non-zero elements which are invariant under Lorentz action.
We are now ready to present the main result of this section.

\begin{theorembox}[$\mathfrak{double}/\mathfrak{string}$ correspondence]
There exists a unique Lorentz-compatible atlas for the Lie $2$-algebra $\mathfrak{string}$ and it consists of the para-K\"{a}hler vector space $\big(\mathbb{R}^{d}\oplus (\mathbb{R}^{d})^\ast,\,J,\,\omega\big)$, where
\begin{itemize}
    \item $J$ is the para-complex structure corresponding to the canonical splitting $\mathbb{R}^{d}\oplus (\mathbb{R}^{d})^\ast$,
    \item $\omega$ is the symplectic structure given by the transgression element of the higher generator of $\mathfrak{string}$ to the space of the atlas $\mathbb{R}^{d}\oplus (\mathbb{R}^{d})^\ast$.
\end{itemize}
\end{theorembox}
\begin{proof}
Recall the definition \ref{def:atlasAlg} of atlas $\phi:\mathfrak{atlas} \longrightarrow \mathfrak{string}$ for an $L_\infty$-algebra. The map $\phi$ can be dually given as an embedding $\phi^\ast:\mathrm{CE}\!\left(\mathfrak{string}\right)\longhookrightarrow \mathrm{CE}(\mathfrak{atlas})$ between their Chevalley-Eilenberg dg-algebras.
Thus, we want to identify an ordinary Lie algebra $\mathfrak{atlas}$ such that its Chevalley-Eilenberg dg-algebra contains a $2$-degree element 
\begin{equation}
    \omega \,:=\, \phi^\ast (B)\;\in\,\mathrm{CE}(\mathfrak{atlas})
\end{equation}
which is the image of the $2$-degree generator of $\mathrm{CE}\!\left(\mathfrak{string}\right)$ and which must satisfy the equation
\begin{equation}
    \di \omega \,=\, 0,
\end{equation}
given by the fact that a homomorphism of dg-algebras maps $\phi^\ast(0)=0$.
Recall that $\mathfrak{atlas}$ must be an ordinary Lie algebra, so its Chevalley-Eilenberg dg-algebra $\mathrm{CE}(\mathfrak{atlas})$ will only have $1$-degree generators. 
Since we want a Lorentz-compatible atlas, $\omega$ must also be a singlet under Lorentz transformations. Thus the generators of the atlas must consist not only in the images $\{e^a:=\phi^\ast(e^a)\}_{a=0,\dots,d-1}$, but also in an extra set $\{\widetilde{e}_a\}_{a=0,\dots,d-1}$ which generates $(\mathbb{R}^d)^\ast$. This way the image of the generator $b\in\mathrm{CE}(\mathfrak{string})$ is
\begin{equation}
    \omega \,=\, \widetilde{e}_a \wedge e^a,
\end{equation}
which is Lorentz-invariant.
Indeed, the generators $\{e^a\}$ of $\mathbb{R}^d$ transform by $e^a\mapsto N^a_{\;b}e^b$ for $N\in SO(1,d-1)$, while the generators $\{\widetilde{e}_a\}$ of $(\mathbb{R}^d)^\ast$ transform by $\widetilde{e}_a\mapsto (N^{-1})^{\;b}_{a}\widetilde{e}_b$.
Now, the equation $\di \omega =0$, combined with the equation $\di e^a = 0$, implies that the differential of the new generator is zero, i.e. $\di \widetilde{e}_a=0$. Therefore, we found the dg-algebra
\begin{equation}
    \mathrm{CE}(\mathfrak{double}) \;=\; \mathbb{R}[e^a,\widetilde{e}_a]/\langle\di e^a=0,\;\di \widetilde{e}_a = 0\rangle
\end{equation}
where we renamed the Lie algebra $\mathfrak{atlas}$ to $\mathfrak{double}$. This ordinary Lie algebra is immediately $\mathfrak{double}=\big(\mathbb{R}^{d}\oplus (\mathbb{R}^{d})^\ast,\,[-,-]=0\big)$, i.e. the abelian Lie algebra whose underlying $2d$-dimensional vector space is $\mathbb{R}^{d}\oplus (\mathbb{R}^{d})^\ast$.
Now, recall that the Chevalley-Eilenberg dg-algebra $\mathrm{CE}(\mathfrak{g})$ of any ordinary Lie algebra $\mathfrak{g}$ is isomorphic to the dg-algebra $\big(\Omega_{\mathrm{li}}^\bullet (G),\,\di\big)$ of left-invariant differential forms on the corresponding Lie group $G=\exp(\mathfrak{g})$. Therefore, we have the isomorphism
\begin{equation}
    \mathrm{CE}(\mathfrak{double}) \;\cong\; \big(\Omega_{\mathrm{li}}^\bullet (\mathbb{R}^{d,d}),\,\di\big)
\end{equation}
where we called $\mathbb{R}^{d,d}$ the abelian Lie group integrating $\mathfrak{double}$ whose underlying smooth manifold is still the linear space $\mathbb{R}^d\times(\mathbb{R}^d)^\ast$. Thus, the smooth functions $\Coo(\mathbb{R}^{d,d})$ are generated by coordinate functions $x^a$ and $\widetilde{x}_a$ and the basis of left-invariant $1$-forms on $\mathbb{R}^{d,d}$ is simply given by
\begin{equation}
    \begin{aligned}
    e^a \,=\, \di x^a,\quad \widetilde{e}_a \,=\, \di \widetilde{x}_a
    \end{aligned}
\end{equation}
Thus, the transgression element $\omega\in\mathrm{CE}(\mathfrak{double})$ is, equivalently, the symplectic form $\omega = \di \widetilde{x}_a \wedge \di x^a$.
Moreover, the canonical splitting $\mathbb{R}^d\oplus(\mathbb{R}^d)^\ast$ induces a canonical para-complex structure $J$, which is compatible with the symplectic form $\omega$. Therefore, the atlas of $\mathfrak{string}$ is equivalently a para-K\"{a}hler vector space $\big(\mathbb{R}^d\oplus(\mathbb{R}^d)^\ast,\,J,\,\omega\big)$.
\end{proof}

\begin{remark}[Emergence of para-Hermitian geometry]
On one side of the correspondence, the Lie $2$-algebra $\mathfrak{string}=\mathbb{R}^d\oplus\mathbf{b}\mathfrak{u}(1)$ is the linearisation of a bundle gerbe and, on the other side, the para-K\"{a}hler vector space $\big(\mathbb{R}^d\oplus(\mathbb{R}^d)^\ast,\,J,\,\omega\big)$ is the linearisation of a para-Hermitian manifold. The latter is an atlas of the former.
\end{remark}

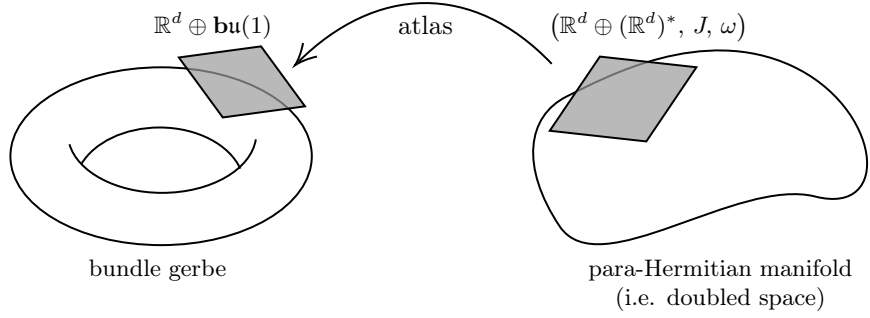
\begin{figure}[h]\begin{center}
\tikzset{every picture/.style={line width=0.75pt}} 
\begin{tikzpicture}[x=0.75pt,y=0.75pt,yscale=-1,xscale=1]
\draw   (10,74.38) .. controls (10,49.59) and (44.03,29.5) .. (86,29.5) .. controls (127.97,29.5) and (162,49.59) .. (162,74.38) .. controls (162,99.16) and (127.97,119.25) .. (86,119.25) .. controls (44.03,119.25) and (10,99.16) .. (10,74.38) -- cycle ;
\draw   (290.5,44.25) .. controls (310.5,34.25) and (359,8) .. (404,28.75) .. controls (449,49.5) and (455.5,104.75) .. (416,94.75) .. controls (376.5,84.75) and (307,139.75) .. (287,109.75) .. controls (267,79.75) and (270.5,54.25) .. (290.5,44.25) -- cycle ;
\draw  [draw opacity=0] (133.77,65.87) .. controls (131.39,81.03) and (111.14,92.87) .. (86.5,92.87) .. controls (63.22,92.87) and (43.84,82.29) .. (39.78,68.33) -- (86.5,62.87) -- cycle ; \draw   (133.77,65.87) .. controls (131.39,81.03) and (111.14,92.87) .. (86.5,92.87) .. controls (63.22,92.87) and (43.84,82.29) .. (39.78,68.33) ;
\draw  [draw opacity=0] (45.78,78.33) .. controls (52.22,67.49) and (67.47,59.88) .. (85.25,59.88) .. controls (104.28,59.88) and (120.4,68.6) .. (125.94,80.65) -- (85.25,89.88) -- cycle ; \draw   (45.78,78.33) .. controls (52.22,67.49) and (67.47,59.88) .. (85.25,59.88) .. controls (104.28,59.88) and (120.4,68.6) .. (125.94,80.65) ;
\draw  [fill={rgb, 255:red, 155; green, 155; blue, 155 }  ,fill opacity=0.7 ] (307.29,24.08) -- (355.99,29.47) -- (330.71,66.92) -- (282.01,61.53) -- cycle ;
\draw  [fill={rgb, 255:red, 155; green, 155; blue, 155 }  ,fill opacity=0.7 ] (136.39,18.93) -- (94.94,24.54) -- (117.12,54.81) -- (158.57,49.2) -- cycle ;
\draw    (155.26,26.71) .. controls (194,-11.83) and (243.11,-14.11) .. (283,27.75) ;
\draw [shift={(153.5,28.5)}, rotate = 314.1] [color={rgb, 255:red, 0; green, 0; blue, 0 }  ][line width=0.75]    (10.93,-4.9) .. controls (6.95,-2.3) and (3.31,-0.67) .. (0,0) .. controls (3.31,0.67) and (6.95,2.3) .. (10.93,4.9)   ;
\draw (204,2) node [anchor=north west][inner sep=0.75pt]  [font=\small] [align=left] {atlas};
\draw (280,0) node [anchor=north west][inner sep=0.75pt]  [font=\footnotesize]  {$\big(\mathbb{R}^{d} \oplus (\mathbb{R}^{d})^{\ast},\, J, \,\omega \big)$};
\draw (81,0) node [anchor=north west][inner sep=0.75pt]  [font=\footnotesize]  {$\mathbb{R}^{d} \oplus \mathbf{b}\mathfrak{u}( 1)$};
\draw (48,125) node [anchor=north west][inner sep=0.75pt]  [font=\footnotesize] [align=left] {bundle gerbe};
\draw (300,125) node [anchor=north west][inner sep=0.75pt]  [font=\footnotesize] [align=left] {\begin{minipage}[lt]{100pt}\setlength\topsep{0pt}
para-Hermitian manifold
\begin{center}
(i.e. doubled space)
\end{center}
\end{minipage}};
\end{tikzpicture}
\caption{para-Hermitian geometry (i.e. the geometry of \\doubled spaces) as atlas description of bundle gerbes.}
\end{center}\vspace{-0.5cm}\end{figure}

\begin{notation}[$\mathfrak{string}$ Lie $2$-algebra]
Usually, the name "string Lie $2$-algebra" is reserved to the central extension $\mathfrak{string}^c(\mathfrak{g})$ of an ordinary Lie algebra $\mathfrak{g}$ by the Lie $2$-algebra $\mathbf{b}\mathfrak{u}(1)$, corresponding to a cocycle $c\in H^3(\mathfrak{g},\mathbb{R})$. However, in this paper, we call $\mathfrak{string}:=\mathbb{R}^{d}\oplus\mathbf{b}\mathfrak{u}(1)$ string Lie $2$-algebra. This has a two-fold reason. Firstly, our definition is a particular case of the latter, even if a trivial one. Secondly, this nomenclature conceptually matches the notation used by \cite{FSS17x, FSS18, FSS18x}, where the symbol "$\mathfrak{string}$" is used to denote the (super) Lie $2$-algebras which can be interpreted as the linearised (super) bundle gerbes underlying Type II and heterotic supergravity.
\end{notation}

\begin{remark}[Kernel pair of the atlas of $\mathfrak{string}$]
Now let us discuss the kernel pair of the atlas $\phi:\mathfrak{atlas} \longrightarrow \mathfrak{string}$. This is defined as the pullback (in the category theory sense) of two copies of the map $\phi$ of the atlas. The homotopy pullback diagram of is
\begin{equation}\label{eq:kernelpair}
\begin{tikzcd}
\mathfrak{double}\times_{\mathfrak{string}}\mathfrak{double} \arrow[r, yshift=0.7ex, two heads] \arrow[r, yshift=-0.7ex, two heads] & \mathfrak{double}  \arrow[r, "\phi"] & \mathfrak{string}.
\end{tikzcd}
\end{equation}
To deal with it, we can consider the Chevalley-Eilenberg algebras of all the involved $L_\infty$-algebras and look at the homotopy pushout diagram of the cokernel pair which is dual to the starting kernel pair \eqref{eq:kernelpair}. This will be given by the following maps of differential graded algebras:
\begin{equation}
\begin{tikzcd}
\mathrm{CE}(\mathfrak{double})\sqcup_{\mathrm{CE}\left(\mathfrak{string}\right)}\mathrm{CE}(\mathfrak{double})  & \mathrm{CE}(\mathfrak{double}) \arrow[l, yshift=0.7ex, hook'] \arrow[l, yshift=-0.7ex, hook']  & \mathrm{CE}(\mathfrak{string}) \arrow[l, hook', "\phi^\ast"'].
\end{tikzcd}
\end{equation}
Let us describe this in more detail. When composed with $\phi^\ast$, the two maps at the centre of the diagram both send the generators $e^a\in \mathrm{CE}(\mathfrak{string})$ to $e^a \in \mathrm{CE}(\mathfrak{double})\sqcup_{\mathrm{CE}\left(\mathfrak{string}\right)}\mathrm{CE}(\mathfrak{double})$. However, they map the generator $B\in \mathrm{CE}(\mathfrak{string})$ to two different elements $\omega=\widetilde{e}_a\wedge e^a$ and $\omega'=\widetilde{e}'_a\wedge e^a$, where $\widetilde{e}_a$ and $\widetilde{e}'_a$ are such that they both satisfy the same equation $\di \widetilde{e}_a' = \di \widetilde{e}_a$. This implies that they are related by a gauge transformation $\widetilde{e}'_a = \widetilde{e}_a + \di\lambda_a$. This can be seen as a consequence of the gauge transformations $B'=B+\di\lambda$ with parameter $\lambda:=\lambda_ae^a$.
\end{remark}

\begin{remark}[T-duality on the $\mathfrak{double}$ algebra]\label{rem:algduality}
The ordinary Lie algebra $\mathfrak{double}$ is not the atlas only of the Lie $2$-algebra $\mathfrak{string}$, but of an entire class of Lie $2$-algebras. For example, we have
\begin{equation}
    \begin{tikzcd}[row sep=5ex, column sep=2ex] 
    & \mathfrak{double} \arrow[dr, "\widetilde{\phi}"]\arrow[dl, "\phi"']& \\
    \mathfrak{string} & & \widetilde{\mathfrak{string}}  
\end{tikzcd}
\end{equation}
where we called $\widetilde{\mathfrak{string}}$ the Lie $2$-algebra whose Chevalley-Eilenberg dg-algebra is given by $\mathrm{CE}\big(\widetilde{\mathfrak{string}}\big)= \mathbb{R}\big[\widetilde{e}_a,\widetilde{B}\big]/\langle\di \widetilde{e}_a = 0,\;\di\widetilde{B}=0\rangle$ and where $\widetilde{\phi}$ is the atlas mapping the generators by $\widetilde{e}_a\mapsto\widetilde{e}_a$ and $\widetilde{B}\mapsto e^a\wedge \widetilde{e}_a$. 
The Lie $2$-algebra $\widetilde{\mathfrak{string}}=(\mathbb{R}^d)^\ast\oplus \mathbf{b}\mathfrak{u}(1)$ can be immediately seen as the T-dualisation of $\mathfrak{string}$ along all the $d$ directions of the underlying spacetime. More generally, $\mathfrak{double}$ will be the atlas of any T-dual of the Lie $2$-algebra $\mathfrak{string}$: this is nothing but a linearised version of T-duality of bundle gerbes.
\end{remark}

\subsection{The doubled space/bundle gerbe correspondence}

In the previous subsection, we established a correspondence between linearised doubled geometries and $L_\infty$-algebras, which interprets the former as an atlas description of the latter. In this subsection we will globalise this relation and we will construct a method to extract a doubled space from a bundle gerbe.

\begin{remark}[On the nature of the extra coordinates]
The $2d$-dimensional atlas of the bundle gerbe is the natural candidate for being an atlas for the doubled space where Double Field Theory lives.
This way, we can completely avoid the conceptual issue of postulating many new extra dimensions in extended geometry, because the extra coordinates which appears in the extended charts describe the degrees of freedom of a bundle gerbe. In this sense, a flat doubled space $\mathbb{R}^{d,d}$ can be seen as a coordinate description of a trivial bundle gerbe. 
\end{remark}

\begin{remark}[Atlas for the Lie $2$-group]
Let $\mathbb{R}^{d}\times \mathbf{B}U(1)$ be the Lie $2$-group which integrates the Lie $2$-algebra $\mathfrak{string}:=\mathbb{R}^{d}\oplus\mathbf{b}\mathfrak{u}(1)$. Let us call again $\mathbb{R}^{d,d}$ the the ordinary Lie group which integrates the ordinary abelian Lie algebra $\mathbb{R}^{d}\oplus (\mathbb{R}^{d})^\ast$. Therefore, we have a homomorphism of Lie groups
\begin{equation}\label{eq:groupatlas}
   \exp(\phi):\,\mathbb{R}^{d,d}\; \longrightarrow\; \mathbb{R}^{d}\times\mathbf{B}U(1),
\end{equation}
which exponentiates the homomorphism of Lie algebras $\phi:\mathfrak{atlas} \longrightarrow \mathfrak{string}$ from the previous section. Consequently, this is also a well defined atlas for $\mathbb{R}^{d}\times\mathbf{B}U(1)$, seen as a smooth stack. 
\end{remark}

\begin{definition}[Lorentz-compatible atlas of an $\infty$-bundle]
Let us define a \textit{Lorentz-compatible atlas} for an $\infty$-bundle $\mathscr{P}\xtwoheadrightarrow{\;\Pi\;} M$ as an atlas whose charts are Lorentz-compatible in the sense of definition \ref{def:lorentzcom}.
\end{definition}

\begin{theoremboxT}[Doubled space from a bundle gerbe]
There exists a unique Lorentz-compatible atlas of a bundle gerbe $\mathscr{G}\xtwoheadrightarrow{\;\Pi\;} M$ with connection and it consists of a para-Hermitian manifold $\left(\M,\,J,\,\omega\right)$, where
\begin{itemize}
    \item $J$ is the para-complex structure corresponding to the splitting of $T\M$ into horizontal and vertical bundle induced by the connection of the bundle gerbe,
    \item $\omega$ is the fundamental $2$-form given by the transgression of the connection of the bundle gerbe, i.e. which satisfies $\pi^\ast H=-\di\omega$ with $H\in\Omega^{3}_{\mathrm{cl}}(M)$ curvature of the bundle gerbe and where $\pi:\mathcal{M}\twoheadrightarrow M$ is the projection induced by $\Pi$ on the atlas.
\end{itemize}
\end{theoremboxT}

\begin{proof}
Let $\mathscr{G}\twoheadrightarrow M$ be a bundle gerbe on a base manifold $M$. Thus $\mathscr{G}$ can be locally trivialised as a collection of local trivial gerbes $\{U_\alpha \times \mathbf{B}U(1)\}_{\alpha\in I}$ on a given open cover $\{U_\alpha\}_{\alpha\in I}$ of the base manifold $M$. 
Thus, we have an effective epimorphism $\varphi_\alpha: \mathbb{R}^{d}\times \mathbf{B}U(1)\rightarrow U_\alpha \times \mathbf{B}U(1)$ for any chart. These can be combined in a single morphism $\bigsqcup_{\alpha\in I}\mathbb{R}^{d}\times \mathbf{B}U(1)\xrightarrow{\;\;\{\varphi_\alpha\}_{\alpha\in I}\;\;} \mathscr{G}$. As explained in \cite{Principal1}, this is in particular an effective epimorphism. To see this, notice that we have a pullback diagram
\begin{equation}
    \begin{tikzcd}[column sep=5ex, row sep=7ex]
    \bigsqcup_{\alpha\in I}\mathbb{R}^{d}\times \mathbf{B}U(1) \arrow[r, two heads]\arrow[d, two heads] & \mathscr{G} \arrow[d, two heads]\\
    \bigsqcup_{\alpha\in I}\mathbb{R}^{d} \arrow[r, two heads] & M,
\end{tikzcd}
\end{equation}
where $\bigsqcup_{\alpha\in I}\mathbb{R}^{d} \twoheadrightarrow M$ is an atlas, and in particular an effective epimorphism, by construction. Since effective epimorphisms are stable under $(\infty,1)$-pullback, the upper arrow is an effective epimorphism too.
Thus, we can cover the bundle gerbe with copies of the Lie $2$-group $\mathbb{R}^d\times\mathbf{B}U(1)$. 
Since this Lie $2$-group comes equipped with the natural atlas \eqref{eq:groupatlas}, we can define the composition maps
$\Phi_\alpha:\mathbb{R}^{d,d} \xrightarrow{\;\;\exp(\phi)\;\;} \mathbb{R}^d\times\mathbf{B}U(1) \xtwoheadrightarrow{\;\;\varphi_\alpha\;\;} U_\alpha \times\mathbf{B}U(1)$. By combining them we can construct an effective epimorphism
\begin{equation}\label{eq:atlas}
    \Phi:\; \bigsqcup_{\alpha\in I}\mathbb{R}^{d,d}\; \xrightarrow{\;\; \{\Phi_\alpha\}_{\alpha\in I} \;\;}\; \mathscr{G} 
\end{equation}
From now on, let us call the total space of the atlas $\M:=\bigsqcup_{\alpha\in I}\mathbb{R}^{d,d}$. Notice that, in general, this is a disjoint union of $\mathbb{R}^{d,d}$-charts.
We can now use the map \eqref{eq:atlas} to explicitly construct the \v{C}ech nerve of the atlas. What we obtain is the following simplicial object:
\begin{equation*}
\begin{tikzcd}[column sep=9ex]
\displaystyle\bigsqcup_{\alpha,\beta,\gamma\in I}\!\!\mathbb{R}^{d,d} \times_\mathscr{G} \mathbb{R}^{d,d} \times_\mathscr{G} \mathbb{R}^{d,d} \arrow[r, yshift=1.4ex, two heads]\arrow[r, two heads] \arrow[r, yshift=-1.4ex, two heads] & \displaystyle\bigsqcup_{\alpha,\beta\in I}\!\mathbb{R}^{d,d} \times_\mathscr{G} \mathbb{R}^{d,d} \arrow[r, yshift=0.7ex, two heads] \arrow[r, yshift=-0.7ex, two heads] & \displaystyle\bigsqcup_{\alpha\in I}\mathbb{R}^{d,d}  \arrow[r, "\{\Phi_\alpha\}_{\alpha\in I} "] & \mathscr{G},
\end{tikzcd}
\end{equation*}
which tells us how the charts of the atlas are glued by morphisms.
Let us describe this diagram in more detail in terms of its dual diagram of function dg-algebras. Let us also call $\big(B_{(\alpha)},\,\Lambda_{(\alpha\beta)},\,G_{(\alpha\beta\gamma)}\big)$ the \v{C}ech cocycle of the bundle gerbe. The two maps of the kernel pair send the local $1$-degree generator to $\di x^\mu$ and the local $2$-degree generator to a couple of local $2$-forms $\omega_{(\alpha)}^{\mathrm{triv}}=\di\widetilde{x}_{(\alpha)\mu}\wedge \di x^\mu$ and $\omega_{(\beta)}^{\mathrm{triv}}=\di\widetilde{x}_{(\beta)\mu}\wedge \di x^\mu$ on the fiber product of the $\alpha$-th and $\beta$-th charts. Now the local $1$-forms $\di\widetilde{x}_{(\alpha)\mu}$ and $\di\widetilde{x}_{(\beta)\mu}$ are required to be related by a gauge transformation $\di\widetilde{x}_{(\alpha)\mu} = \di\widetilde{x}_{(\beta)\mu} + \di\Lambda_{(\alpha\beta)\mu}$ where the gauge parameters $\Lambda_{(\alpha\beta)\mu}$ are given by the cocycle of the bundle gerbe. Equivalently, the two $2$-forms must be related by a gauge transformation $\omega_{(\alpha)}^{\mathrm{triv}}=\omega_{(\beta)}^{\mathrm{triv}}+\di\Lambda_{(\alpha\beta)}$  with gauge parameter $\Lambda_{(\alpha\beta)}:=\Lambda_{(\alpha\beta)\mu}\di x^\mu$. The gauge parameters are required to satisfy the cocycle condition $\Lambda_{(\alpha\beta)}+\Lambda_{(\beta\gamma)}+\Lambda_{(\gamma\alpha)}=\di G_{(\alpha\beta\gamma)}$ on three-fold overlaps of charts.

\noindent On the atlas $\M$ of the bundle gerbe, we can define a $2$-form $\omega\in\Omega^2(\M)$ by taking the difference $\omega_{(\alpha)} := \omega^{\mathrm{triv}}_{(\alpha)} - \pi^\ast B_{(\alpha)}$ of the local $2$-form $\omega^{\mathrm{triv}}_{(\alpha)}$ and the pullback of the local connection $2$-form $B_{(\alpha)}$ of the bundle gerbe from the base manifold on each chart $\mathbb{R}^{d,d}$. This definition assures that $\omega_{(\alpha)} =  \omega_{(\beta)}$ on overlaps of charts $\mathbb{R}^{d,d}\times_\mathscr{G}\mathbb{R}^{d,d}$. Therefore, this $2$-form is globally well-defined and we can write it simply as $\omega$, by removing the $\alpha$-index. In local coordinates we can write
\begin{equation}\label{eq:omegarev}
    \omega \;=\; \big(\di\widetilde{x}_{(\alpha)\mu}+B_{(\alpha)\mu\nu}\di x^\nu \big)\wedge \di x^\mu
\end{equation}
Notice that the form $\omega$ is, more generally, invariant under gauge transformations of the bundle gerbe. From the definition of $\omega$, we obtain the relation with curvature of the bundle gerbe:
\begin{equation}
    \pi^\ast H\,=\,-\di\omega, \quad \text{with} \quad H\in\Omega^3_{\mathrm{cl}}(M),
\end{equation}
where $H\in\Omega^3_\mathrm{cl}(M)$ is the curvature of the bundle gerbe, i.e. a closed $3$-form which satisfies the equation $H|_{U_\alpha}=\di B_{(\alpha)}$ on any open set $U_\alpha\subset M$ of our cover.
Now, we want to show that $\M$ is canonically para-Hermitian with fundamental $2$-form $\omega$. The projection $\pi: \M \twoheadrightarrow M$ induces a short exact sequence of vector bundles: 
\begin{equation}\label{eq1}
    0 \longhookrightarrow \mathrm{Ker}(\pi_\ast) \longhookrightarrow T\M \xtwoheadrightarrow{\;\pi_\ast\;} \pi^\ast TM \xtwoheadrightarrow{\quad} 0.
\end{equation}
The fundamental $2$-form $\omega$ immediately induces a map $\omega^\sharp:T\mathcal{M}\rightarrow T\mathcal{M}$. In particular, this is a projector to the vertical bundle, i.e.
\begin{equation}
    \omega^\sharp:\, T\M \,\longtwoheadrightarrow\, \mathrm{Ker}(\pi_\ast).
\end{equation}
To see this, it is enough to notice that the fundamental $2$-form $\omega = (\di \widetilde{x}_{(\alpha)\mu} + B_{(\alpha)\mu\nu}\di x^\nu)\wedge \di x^\mu$ induces the map $\omega^\sharp(V) = (\widetilde{v}_{\mu} + B_{(\alpha)\mu\nu} v^\nu)\widetilde{\partial}^\mu\in\mathrm{Ker}(\pi_\ast)$, where $V = v^\mu\partial_\mu + \widetilde{v}_{\mu}\widetilde{\partial}^\mu$ is any vector on the atlas, expressed in the coordinate basis.
Therefore, the $2$-form $\omega$ defines the splitting $\pi_\ast\oplus\omega^\sharp$ into horizontal and vertical bundle
\begin{equation}\label{eq2}
    T\M \; \cong\; \pi^\ast TM \,\oplus\,  \mathrm{Ker}(\pi_\ast).
\end{equation}
This splitting canonically defines a para-complex structure $J\in\mathrm{Aut}(T\M)$. If we split any vector in horizontal and vertical projection $X=X_H+X_V$, the para-complex structure $J$ is defined such that $J(X)= X_H-X_V$. Notice that, since $J$ defines a splitting $T\M =L_+ \oplus L_-$ of the tangent bundle of $\M$, as seen in section \ref{para}, this identifies $L_+ \equiv \pi^\ast TM$ and $L_-\equiv\mathrm{Ker}(\pi_\ast)$.
Therefore, the atlas of a bundle gerbe is a para-Hermitian manifold $(\M,\,J,\,\omega)$ with para-complex structure $J$ and fundamental $2$-form $\omega$, defined above.
\end{proof}

\begin{remark}[Towards an extended/higher correspondence]
We can easily notice that not any para-Hermitian manifold can be obtained as an atlas of a bundle gerbe with connection. Therefore, there exists no bijection between para-Hermitian geometries and bundle gerbes with connection. 
However, as we are going to see in the final section of this paper, the para-Hermitian atlas of a given bundle gerbe is also an atlas for any string background which is T-dual to the original bundle gerbe.
As shown in \cite{Alf19, Alf20}, most of the known examples of doubled spaces, including the ones underlying non-geometric or non-abelian T-duality, can be derived from the structure of a bundle gerbe with connection. 
Therefore, our method of extracting a para-Hermitian geometry from a bundle gerbe can be applied to all the relevant examples of doubled spaces coming from the physics literature.
Moreover, the idea that a general doubled space must be an (almost) para-Hermitian manifold is something which has been postulated, not strictly proved.
Thus, it is legitimate to speculate towards the establishment of a full extended/higher correspondence.
\end{remark}

\begin{remark}[Principal connection of the bundle gerbe]\label{remarkrev}
Let $\Omega^2\in\mathbf{H}$ be the usual sheaf of differential $2$-forms over smooth manifolds.
We can define a differential $2$-form $\underline{\omega}$ on the bundle gerbe $\mathscr{G}$ as a map  $\mathscr{G}\xrightarrow{\;\underline{\omega}\;}\Omega^2$. Notice that, given the fundamental $2$-form $\omega$ in \eqref{eq:omegarev}, we can construct a $2$-form $\mathscr{G}\xrightarrow{\;\underline{\omega}\;}\Omega^2$ given as follows:
\begin{equation*}
\begin{tikzcd}[column sep=7ex, row sep=7ex]
\M \arrow[r, "\Phi_{\alpha}"]\arrow[rr, bend right=35, "\omega_{(\alpha)}"]& \mathscr{G} \arrow[r, "\underline{\omega}"] & \Omega^2
\end{tikzcd}, \qquad
\begin{tikzcd}[column sep=7ex, row sep=7ex]
\M \times_\mathscr{G} \M \arrow[bend left=40, "\omega_{(\alpha)}"]{r}[name=U,below]{}
\arrow[bend right=40, "\omega_{(\beta)}"']{r}[name=D]{} &
\Omega^2 \arrow[Rightarrow, to path=(U) -- (D)]{}
\end{tikzcd}.
\end{equation*}
Since $\Omega^2$ is a $0$-truncated stack, the $2$-morphism in the second diagram is just an identity. In other words we obtain that $\underline{\omega}\in\Omega^2(\mathscr{G})$ is given on the atlas by a collection of local $2$-forms $\omega_{(\alpha)} =\big(\di\widetilde{x}_{(\alpha)\mu}+B_{(\alpha)\mu\nu}\di x^\nu \big)\wedge \di x^\mu$ on any chart, which satisfy $\omega_{(\alpha)}=\omega_{(\beta)}$ on any overlap of charts. Thus, the fundamental $2$-form $\omega$ on the atlas $\mathcal{M}$ from the previous theorem can be interpreted as a $2$-form $\underline{\omega}$ on the bundle gerbe $\mathscr{G}$.
\end{remark}

\begin{remark}[Analogy with a principal $U(1)$-bundle]
The way of expressing the fundamental $2$-form on our atlas as in remark \ref{remarkrev} is, despite of the appearance, very natural and familiar. When we write the connection of a $U(1)$-bundle in local coordinates, we are exactly writing a $1$-form $\omega_{(\alpha)} := \di\theta_{(\alpha)}+A_{(\alpha)\mu}(x_{(\alpha)})\di x^\mu_{(\alpha)}\in\Omega^1(\mathbb{R}^{d+1})$ on the local chart $\mathbb{R}^{d+1}$, where $\big\{\di\theta_{(\alpha)},\di x^\mu_{(\alpha)}\big\}$ is the coordinate basis of $\Omega^1(\mathbb{R}^{d+1})$. On the overlaps of charts we have $\omega_{(\alpha)}=\omega_{(\beta)}$, which assures that the the $1$-form we are writing in local coordinates is equivalently the pullback $\omega_{(\alpha)}=\phi_\alpha^\ast\underline{\omega}$ of a well-defined $1$-form $\underline{\omega}$ on the total space of the $U(1)$-bundle.
Notice that this is in perfect analogy with remark \ref{remarkrev}.
\end{remark}

\begin{example}[Topologically trivial doubled space]
Let us consider a topologically trivial bundle gerbe $\mathscr{G}=M\times \mathbf{B}U(1)$ with connection.
The corresponding doubled space is a para-K\"{a}hler manifold $(\M,J,\omega)$ where $\M= T^\ast M$ is just the cotangent bundle of the base manifold, the para-complex structure $J$ corresponds to the canonical splitting $T\M \cong TM\oplus T^\ast M$ and the connection $\omega = \di\widetilde{x}_\mu\wedge \di x^\mu$ is the canonical symplectic form on $T^\ast M$ with $\{x^\mu,\widetilde{x}_\mu\}$ Darboux coordinates.
\end{example}

\begin{example}[Doubled Minkowski space]
If, in the previous example, we choose as base manifold the Minkowski space $M=\mathbb{R}^d$, the corresponding doubled space will be the para-K\"{a}hler vector space $(\mathbb{R}^{d,d},J,\omega)$.
\end{example}

\begin{remark}[Correspondence between sections of the bundle gerbe and the doubled space]
Let us consider again a topologically trivial bundle gerbe $\mathscr{G}=M\times \mathbf{B}U(1)$ with connection. Any section $M\xhookrightarrow{\;I\;}\mathscr{G}$ will be a $U(1)$-bundle $I\twoheadrightarrow M$, while any section $M\xhookrightarrow{\;\iota\;}\M$ will be an embedding $\widetilde{x}=\widetilde{x}(x)$. These two objects are immediately related by
\begin{equation}
    \iota^\ast\omega \;=\; \mathrm{curv}(I)
\end{equation}
where $\mathrm{curv}(-)$ is the curvature $2$-form of a $U(1)$-bundle. Since any bundle gerbe can be locally trivialised, it is possible to generalise this relation to the general topologically non-trivial case. The relation between bundle gerbes and doubled spaces was firstly presented and studied in \cite{Alf19} by using this observation.
\end{remark}

\begin{remark}[A doubled-yet-gauged space]\label{d-y-g}
The principal action of the bundle gerbe is transgressed to the atlas by a shift $(x^\mu,\widetilde{x}_\mu)\mapsto(x^\mu,\widetilde{x}_\mu+\lambda_\mu(x))$ in the unphysical coordinates, which can be identified with a gauge transformation $B\mapsto B + \di(\lambda_\mu\di x^\mu)$ of the Kalb-Ramond field.
Moreover, the property $\mathscr{G}/\mathbf{B}U(1)\cong M$ of bundle gerbes, when transgressed to the atlas, can be identified with the idea that physical points correspond to gauge orbits of the doubled space \cite{Park13}. Remarkably, this gives a global geometric interpretation of the strong constraint of Double Field Theory \cite{Alf19}. 
Therefore, an atlas of the bundle gerbe is naturally a doubled-yet-gauged space, according to the definition given by \cite{Park13}.
\end{remark}

\begin{remark}[Basis of global forms]
In general it is also possible to express the principal connection $\omega=\widetilde{e}_{a}\wedge e^a$ in terms of the globally defined $1$-forms $\widetilde{e}_{a}=\di\widetilde{x}_{(\alpha)a} + B_{(\alpha)a\nu}\di x^\nu$ and $e^a=\di x^a$ on the atlas.
We pack both in a single global $1$-form $E^A$ with index $A=1,\dots,2d$ which is defined by $E^a:=e^a$ and $E_{a}:=\widetilde{e}_{a}$. In this basis, we have that the connection can be expressed by $\omega \;=\; \omega_{AB}\,E^A\wedge E^B$, where $\omega_{AB}$ is the $2d$-dimensional standard symplectic matrix.
\end{remark}

\begin{definition}[Generalised metric]
A global generalised metric can be defined, in analogy with a Riemannian metric, as an orthogonal structure $\mathscr{G}\xrightarrow{\;\underline{\mathcal{G}}\;}O(2d)\mathrm{Struc}$ on the bundle gerbe, where the stack $O(2d)\mathrm{Struc}$ which we will now construct explicitly. A map $\mathscr{G}\rightarrow O(2d)\mathrm{Struc}$ is defined as a twisted bundle given as it follows:
\begin{equation}
\begin{tikzcd}[column sep=6ex, row sep=7ex]
T\mathscr{G} \arrow[r]\arrow[d, two heads] & \arrow[r]\arrow[d, two heads]  GL(2d)/\!/O(2d) & \arrow[d, two heads] \ast \\
\mathscr{G} \arrow[r]\arrow[rr, bend right, "\mathcal{N}_{(\alpha\beta)}"] & \mathbf{B}O(2d) \arrow[r]  & \mathbf{B}GL(2d).
\end{tikzcd}
\end{equation}
The bundle $T\mathscr{G}$ is naturally classified by a cocycle valued in $\mathbf{B}GL(d)\ltimes\mathbf{b}\mathfrak{u}(1)_\mathrm{conn}$ on the base manifold $M$ of the bundle gerbe. Notice that this can be embedded in a cocycle valued in $\mathbf{B}O(d,d)$, i.e.
\begin{equation}
     \mathbf{B}GL(d)\ltimes\mathbf{b}\mathfrak{u}(1)_\mathrm{conn} \;\hookrightarrow\;\mathbf{B}O(d,d)\;\hookrightarrow\;\mathbf{B}GL(2d).
\end{equation}
Such a cocycle is given by the following $O(d,d)$-valued matrices on each overlap of patches:
\begin{equation}
    \mathcal{N}_{(\alpha\beta)} \;=\;   \begin{pmatrix}
 N_{(\alpha\beta)} & 0 \\
 \mathrm{d}\Lambda_{(\alpha\beta)} & N_{(\alpha\beta)}^{-\mathrm{T}}
 \end{pmatrix},
\end{equation}
where $N_{(\alpha\beta)}$ are the transition functions corresponding to $TM$ and $(\Lambda_{(\alpha\beta)},G_{(\alpha\beta\gamma)})$ is the \v{C}ech cocycle corresponding to the bundle gerbe.
The cocycle $\mathcal{N}_{(\alpha\beta)}$ can be seen as the cocycle corresponding to $T\M$ appearing in the short exact sequence \eqref{eq1}. 
Moreover, notice that $O(d,d)\cap O(2d)\cong O(d)\times O(d)$. Therefore the inclusion $\mathbf{B}O(2d)\hookrightarrow \mathbf{B}GL(2d)$ which defines a general orthogonal structure reduces to $\mathbf{B}\big(O(d)\times O(d)\big)\hookrightarrow \mathbf{B}O(d,d)$.

\noindent On the atlas $(\M,J,\omega)$, this will be given by a collection of Riemannian metrics $\mathcal{G}_{(\alpha)}$ with the following patching conditions:
\begin{equation*}
\begin{tikzcd}[column sep=7ex, row sep=7ex]
\M \arrow[r, "\Phi_{\alpha}"]\arrow[rr, bend right=35, "\mathcal{G}_{(\alpha)}"]& \mathscr{G} \arrow[r, "\underline{\mathcal{G}}"] &  O(2d)\mathrm{Struc}.
\end{tikzcd}, \qquad
\begin{tikzcd}[column sep=7ex, row sep=7ex]
\M \times_\mathscr{G} \M \arrow[bend left=40, "\mathcal{G}_{(\alpha)}"]{r}[name=U,below]{}
\arrow[bend right=40, "\mathcal{G}_{(\beta)}"']{r}[name=D]{} &
 O(2d)\mathrm{Struc}, \arrow[Rightarrow, to path=(U) -- (D)]{}
\end{tikzcd}
\end{equation*}
which assures that they are patched by the condition $\mathcal{G}_{(\beta)}=\mathcal{N}_{(\alpha\beta)}^{\mathrm{T}}\mathcal{G}_{(\alpha)}\mathcal{N}_{(\alpha\beta)}$.
\end{definition}

\noindent As explained in \cite{Alf19}, if we require the generalised metric structure to be invariant under the principal $\mathbf{B}U(1)$-action of the bundle gerbe, this will have to be of the form
\begin{equation}
\begin{aligned}
    \mathcal{G} \;=\; \mathcal{G}_{AB}\,E^A \odot E^B \;=\; g_{ab}\,e^a\odot e^b \,+\, g^{ab}\,\widetilde{e}_{a} \odot \widetilde{e}_{b} 
    \end{aligned}
\end{equation}
where we called the matrix $\mathcal{G}_{AB}:=(g\oplus g^{-1})_{AB}$ and where $g\in\odot^2\Omega^1(M)$ is a Riemannian metric on the base manifold. In the coordinate basis $\{\di x^\mu_{(\alpha)},\di\widetilde{x}_{(\alpha)\mu}\}$ we find the usual expression
\begin{equation}
    \mathcal{G}_{(\alpha)MN} \;=\; \begin{pmatrix}g_{\mu\nu}- B_{(\alpha)\mu\lambda}g^{\lambda\rho}B_{(\alpha)\rho\beta} & B_{(\alpha)\mu\lambda}g^{\lambda\nu} \\-g^{\mu\lambda}B_{(\alpha)\lambda\nu} & g^{\mu\nu} \end{pmatrix},
\end{equation}
where $B_{(\alpha)}$ is the connection of the bundle gerbe. As explained in remark \ref{d-y-g}, invariance under the principal $\mathbf{B}U(1)_{\mathrm{conn}}$-action can be seen as the global geometric version of the strong constraint of Double Field Theory. This was called higher cylindricity condition in \cite{Alf19}, in analogy with Kaluza-Klein theory.

\subsection{The NS5-brane is a higher Kaluza-Klein monopole}

In this subsection we will present an immediate application of the correspondence between doubled spaces and bundle gerbes. We will, indeed, formalise the NS5-brane of $10$-dimensional supergravity as a topologically non-trivial higher Kaluza-Klein monopole on the bundle gerbe.

\begin{definition}[Higher Dirac monopole]\label{def:hdm}
A higher Dirac monopole is a topologically non-trivial bundle gerbe $\mathscr{G}\longtwoheadrightarrow \mathbb{R}^{1,5} \times \left(\mathbb{R}^4-\{0\}\right)$.
\end{definition}

\noindent Here, $\mathbb{R}^4-\{0\}$ can be seen as the transverse space of the monopole and $\mathbb{R}^{1,5}$ as its world-volume, magnetically charged by the Kalb-Ramond field.

\begin{remark}[Higher Dirac charge-quantization]
Notice that $\mathbb{R}^4-\{0\}\simeq\mathbb{R}^+ \times S^3$, where $\mathbb{R}^+$ gives the radial direction in the transverse space and $S^3$ the angular directions. Since $\mathbb{R}^{1,5} \times \mathbb{R}^+ \times S^3$ is homotopy equivalent to $S^3$, we have $\mathrm{dd}(\mathscr{G})\in H^3(S^3,\mathbb{Z})\cong\mathbb{Z}$. This implies
\begin{equation}
    \mathrm{dd}(\mathscr{G}) = \frac{m}{2}\,[\mathrm{Vol}(S^3)],
\end{equation}
with $m\in\mathbb{Z}$, in direct analogy with the ordinary Dirac monopole.
\end{remark}

\noindent Now we can give a precise definition of a higher Kaluza-Klein monopole, which is constructed by directly generalising the ordinary Kaluza-Klein monopole \cite{GrosPer83} to a bundle gerbe.

\begin{definition}[Higher Kaluza-Klein monopole]\label{def:hkkm}
A higher Kaluza-Klein monopole \cite{Alf19} is a non-trivial bundle gerbe $\mathscr{G}\longtwoheadrightarrow \mathbb{R}^{1,5} \times \left(\mathbb{R}^4-\{0\}\right)$ equipped with a generalised metric $\mathcal{G}$ such that, on the atlas $\M$, it takes the form
\begin{equation}\label{eq:hkkmonopole}
    \begin{aligned}
        \mathcal{G} &= \eta_{\mu\nu}\mathrm{d}x^\mu\mathrm{d}x^\nu + \eta^{\mu\nu}\mathrm{d}\widetilde{x}_\mu\mathrm{d}\widetilde{x}_\nu + h(r)\delta_{ij}\mathrm{d}y^i\mathrm{d}y^j + \frac{\delta^{ij}}{h(r)}(\mathrm{d}\widetilde{y}_i+B_{ik}\mathrm{d}y^k)(\mathrm{d}\widetilde{y}_j+B_{jk}\mathrm{d}y^k)
    \end{aligned}
\end{equation}
where the curvature of the gerbe and the harmonic function are respectively
\begin{equation}\label{eq:transversalcond}
    H = \star_{\mathbb{R}^4}\mathrm{d}h, \quad h(r)=1+\frac{m}{r^2}
\end{equation}
with $m\in\mathbb{Z}$ and $r^2:=\delta_{ij}y^iy^j$ radius in the four dimensional transverse space.
Here, the atlas $(\M,\omega,J)$ of the bundle gerbe, with fundamental $2$-form $\omega = \mathrm{d}\widetilde{x}_\mu\wedge\di x^\mu + (\mathrm{d}\widetilde{y}_i + B_{ij}\di y^j) \wedge \di y^i $ and $\{x^\mu, \widetilde{x}_\mu\}$ are coordinates on $T^\ast \mathbb{R}^{1,5}$ and $\{y^i, \widetilde{y}_i\}$ are local coordinates on $\M|_{\mathbb{R}^4-\{0\}}$.
\end{definition}

\noindent Notice that this monopole is nothing but a globally-defined Berman-Rudolph monopole \cite{BR14}. As observed by \cite{Bakhmatov:2016kfn}, the Berman-Rudolph monopole gives rise to the non-geometric branes. In the global geometric context, the arising of non-geometric branes was studied in \cite{Alf19}.

\begin{remark}[NS5-brane is higher Kaluza-Klein monopole]
By higher Kaluza-Klein reduction of \eqref{eq:hkkmonopole} to $M=\mathbb{R}^{1,5}\times\mathbb{R}^+\times S^3$ we get the following metric and gerbe connection
\begin{equation}
    \begin{aligned}
        g = \eta_{\mu\nu}\mathrm{d}x^\mu\mathrm{d}x^\nu + h(r)\delta_{ij}\mathrm{d}y^i\mathrm{d}y^j,\qquad B= B_{ij}\,\mathrm{d}y^i\wedge\mathrm{d}y^j
    \end{aligned}
\end{equation}
which satisfy the conditions \eqref{eq:transversalcond} on the transverse space.
These are exactly the metric and Kalb-Ramond field of an NS5-brane with $H$-charge $m\in\mathbb{Z}$ in $10d$ spacetime $M$.
\end{remark}

\noindent Therefore, the higher Kaluza-Klein monopole encompasses a higher Dirac monopole from definition \ref{def:hdm}, just as the Kaluza-Klein monopole does with an ordinary Dirac monopole. 
The Kaluza-Klein brane appears when spacetime is a non-trivial circle bundle and, analogously, the NS5-brane appears when the bundle gerbe is non-trivial.

\subsection{Recovering generalised geometry}

Here we will show that generalised geometry is naturally recovered from the bundle gerbe perspective upon imposition of the strong constraint, i.e. invariance under the principal $\mathbf{B}U(1)$-action.

\begin{remark}[Generalised geometry on the atlas]\label{vecatlas}
Let $\{\partial_M\}=\{\partial_\mu, \widetilde{\partial}^\mu\}$ be the local coordinate basis of $T\M$.
A vector on the atlas $(\M,J,\omega)$ can be written in local coordinates as $V=V^M_{(\alpha)}\partial_M = v^\mu_{(\alpha)} \partial_\mu + \widetilde{v}_{(\alpha)\mu}\widetilde{\partial}^\mu$, where the components $V^M_{(\alpha)}$ are locally defined. The fundamental $2$-form $\omega$ will project this into a vertical vector $\omega(V)=(\widetilde{v}_{(\alpha)\mu}+B_{(\alpha)\mu\nu}v^\nu_{(\alpha)})\widetilde{\partial}^\mu$. 
Now, if we call $\{D_A\}$ the basis of globally defined vectors on $\M$ dual to the global $1$-forms $\{E^A\}$, we can write a vector on the atlas by $V=V^AD_A$, where now the components $V^A$ are globally defined.
We can now express the isomorphism $\pi_\ast\oplus\omega$ in \eqref{eq2} by
\begin{equation}
    V^AD_A\;=\; v^\mu_{(\alpha)} \partial_\mu + \big(\widetilde{v}_{{(\alpha)}\mu}+B_{(\alpha)\mu\nu}v^\nu_{(\alpha)}\big)\widetilde{\partial}^\mu
\end{equation}
Notice that, if we restrict ourselves to strong constrained vectors, i.e. vectors whose components $V^M_{(\alpha)}$ only depend on the coordinates of the base manifold $M$, these are immediately sections of a Courant algebroid twisted by the bundle gerbe $\mathscr{G}\twoheadrightarrow M$ with local potential $B_{(\alpha)}$.
\end{remark}
 
\noindent We have already shown that strong constrained vectors on such an atlas reduce to sections of a Courant algebroid. Now, we want to show that the bracket structure of the Courant algebroid also comes from the bundle gerbe. This was mostly explored in \cite{Alf19}. \vspace{0.2cm}

\noindent We can now introduce the infinitesimally thickened point
\begin{equation}
    \mathbb{D}^1 \;:=\; \mathrm{Spec}(\mathbb{R}[\epsilon]/\langle\epsilon^2\rangle).
\end{equation}
Notice that this is not a stack, i.e. $\mathbb{D}^1 \notin \mathbf{H}$. However, it is possible to define a new $(\infty,1)$-category $\mathbf{H}_{\mathrm{formal}}$ by enlarging the category $\mathbf{Diff}$ of smooth manifolds, on which the objects of $\mathbf{H}$ are presheaves. It well-understood that this can be achieved by using the category $\mathbf{Diff}_\mathrm{formal}$ of formal smooth manifolds, i.e. smooth manifolds possibly equipped with infinitesimal extension. For details about this construction see \cite{DCCTv2, khavkine2017synthetic}. From now on, we will commit a slight abuse of notation and we will denote $\mathbf{H}_{\mathrm{formal}}$ just by $\mathbf{H}$.

\begin{definition}[Tangent stack]
We define the tangent stack of a stack $\mathscr{X}\in\mathbf{H}$ as the internal hom stack $T\mathscr{X} := [\mathbb{D}^1,\mathscr{X}]$, where $\mathbb{D}^1$ is the infinitesimally thickened point.
\end{definition}

\begin{remark}[Atiyah sequence of the bundle gerbe]
For a given a bundle gerbe with connective structure $\mathscr{G}\xtwoheadrightarrow{\;\pi\;}M$, we can define its tangent stack by $T\mathscr{G} = [\mathbb{D}^1,\mathscr{G}]$. 
A direct calculation shows that $[\mathbb{D}^1,M]=TM$ and $[\mathbb{D}^1,\mathbf{B}U(1)_{\mathrm{conn}}]=\mathbf{B}U(1)_{\mathrm{conn}}\ltimes \mathbf{b}\mathfrak{u}(1)_{\mathrm{conn}}$, where $\mathbf{b}\mathfrak{u}(1)_{\mathrm{conn}}$ is the stack of real line bundles with connection.
From this, we obtain the short exact sequence
\begin{equation}
    0 \longhookrightarrow \mathscr{G}\ltimes\mathbf{b}\mathfrak{u}(1)_{\mathrm{conn}} \longhookrightarrow T\mathscr{G} \xtwoheadrightarrow{\;\pi_\ast\;} \pi^\ast TM \xtwoheadrightarrow{\quad} 0.
\end{equation}
This sequence is nothing but the stack version of the short exact sequence \eqref{eq1}.
If we also choose a connection $B_{(\alpha)}$ for the bundle gerbe, we will have induced an isomorphism of stacks
\begin{equation}\label{eqa}
    T\mathscr{G} \; \cong\; \pi^\ast TM \,\oplus\, \mathscr{G}\ltimes\mathbf{b}\mathfrak{u}(1)_{\mathrm{conn}},
\end{equation}
which is the stack version of the isomorphism \eqref{eq2}.
\end{remark}

\begin{definition}[Atiyah $L_\infty$-algebroid of the bundle gerbe]
We can define the Atiyah $L_\infty$-algebroid of our bundle gerbe by $\mathfrak{at}_\mathscr{G} := T\mathscr{G}/\!/\mathbf{B}U(1)_{\mathrm{conn}} \longtwoheadrightarrow M$, in perfect analogy with the Atiyah algebroid of a principal bundle.
\end{definition}

\begin{remark}[Courant $2$-algebra]
The isomorphism \eqref{eqa} induces the isomorphism of $L_\infty$-algebroids on the manifold $M$
\begin{equation}
    \mathfrak{at}_\mathscr{G} \; \cong\; TM \,\oplus_\mathrm{s}\, M\times\mathbf{b}\mathfrak{u}(1)_{\mathrm{conn}},
\end{equation}
where $\oplus_\mathrm{s}$ is the semi-direct sum.
This, on sections, gives the isomorphism of $L_\infty$-algebras
\begin{equation}
    \Gamma(M,\mathfrak{at}_\mathscr{G}) \; \cong\; \mathfrak{X}(M) \,\oplus_\mathrm{s}\, \mathbf{b}\mathfrak{u}(1)_{\mathrm{conn}}(M),
\end{equation}
where $\mathbf{b}\mathfrak{u}(1)_{\mathrm{conn}}(M)$ is the $2$-algebra of line $\mathfrak{u}(1)$-bundles with connection on $M$.
In \cite{Alf19} we showed that the sections of such algebroid encode the infinitesimal symmetries of a bundle gerbe with connective structure.
As also seen in \cite{Alf19}, a section $V\in\Gamma(M,\mathfrak{at}_\mathscr{G})$ can be expressed in \v{C}ech data as $V=(v+\widetilde{v}_{(\alpha)}, \, f_{(\alpha\beta)})$, where $v\in\mathfrak{X}(M)$ is a global vector field, $\widetilde{v}_{(\alpha)}\in\Omega^1(U_\alpha)$ is a collection of $1$-forms on each patch $U_\alpha$ of $M$ and $f_{(\alpha\beta)}\in\Coo(U_\alpha\cap U_\beta)$ is a collection of functions on each overlap $U_\alpha\cap U_\beta$ of $M$. 
These local data are glued according to
\begin{equation}\label{eq:pathcthevector}
\begin{aligned}
    \begin{aligned}
        \xi_{(\alpha)} - \xi_{(\beta)} \;&=\; -\iota_{X}\mathrm{d}\Lambda_{(\alpha\beta)}+\mathrm{d}f_{(\alpha\beta)}, \\
        f_{(\alpha\beta)}+f_{(\beta\gamma)}+f_{(\gamma\alpha)} \;&=\; 0,
    \end{aligned}
\end{aligned}
\end{equation}
where $(\Lambda_{(\alpha\beta)},G_{(\alpha\beta\gamma)})$ is the \v{C}ech-Deligne cocycle corresponding to the connective structure of the bundle gerbe $\mathscr{G}$. \vspace{0.25cm}

\noindent As shown by \cite{Col11}, the Lie $2$-algebra structure of $\mathfrak{X}(M)\oplus_\mathrm{s} \mathbf{b}\mathfrak{u}(1)_{\mathrm{conn}}(M)$ is isomorphic to the Lie $2$-algebra structure of the standard Courant $2$-algebra, whose $2$-bracket is the Courant bracket $[ -, -]_{\mathrm{Cou}}$.
If we write sections $V,W\in \Gamma(M,\mathfrak{at}_\mathscr{G})$ of the Atiyah $L_\infty$-algebroid on the atlas, in the notation of remark \ref{vecatlas}, we will have the $2$-bracket
\begin{equation}
    [ V, W ]_{\mathrm{Cou}} \;=\;\, [v,w]_{\mathrm{Lie}} + \mathcal{L}_v\widetilde{w} - \mathcal{L}_w\widetilde{v} - \frac{1}{2}\di(\iota_v\widetilde{w}-\iota_w\widetilde{v}) + \iota_v\iota_w H,
\end{equation}
where $H\in\Omega^3_\mathrm{cl}(M)$ is the curvature of the gerbe.
\end{remark} 

\noindent Let us conclude this section by mentioning the relation between this stack perspective on generalised geometry and symplectic dg-geometry.

\begin{remark}[Relation with NQP-manifolds]
It is well-understood that, given a $L_\infty$-algebroid $\mathfrak{a}\twoheadrightarrow M$, its Chevalley-Eilenberg dg-algebra $\mathrm{CE}(\mathfrak{a})$ can be seen as the dg-algebra of functions on a dg-manifold, also known as NQ-manifold. In the case of the Atiyah $L_\infty$-algebroid, we have
\begin{equation}
    \mathrm{CE}(\mathfrak{at}_\mathscr{G}) \; =\; \Big(  \Coo(T^\ast[2]T[1]M),\,Q_H\Big)
\end{equation}
where the dg-manifold $T^\ast[2]T[1]M$, called Vinogradov algebroid, is canonically symplectic, i.e. it is a NQP-manifold. The Poisson bracket, combined with the differential $Q_H$, reproduces the Courant $2$-algebra \cite{Roy02}.
Inspired by this relation, a purely dg-geometric approach to Double Field Theory was developed by \cite{DesSae18, DesSae18x, Crow-Watamura:2018liw, DesSae19}.
\end{remark}

\section{T-duality, non-geometry and bundle gerbes}\label{s5}

Recall that we already described a linearised version of T-duality in remark \ref{rem:algduality}, where we showed that every couple of T-dual Lie $2$-algebras share the same atlas. \vspace{0.2cm}

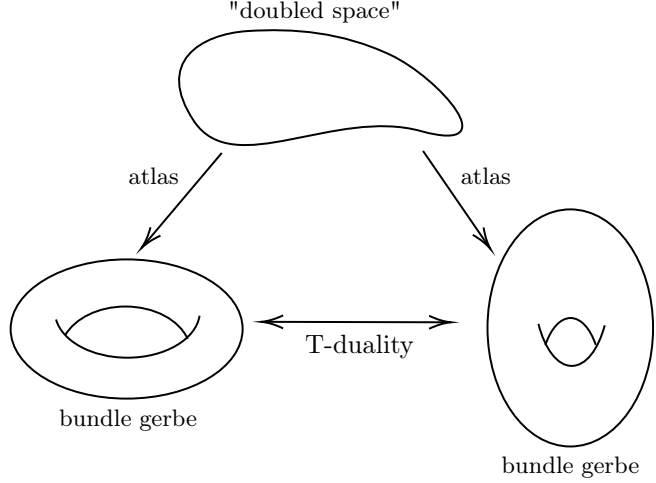
\begin{figure}[h]\begin{center}
\tikzset{every picture/.style={line width=0.75pt}} 
\begin{tikzpicture}[x=0.75pt,y=0.75pt,yscale=-1,xscale=1]
\draw   (123,3.25) .. controls (146.5,1.75) and (170.5,3.75) .. (195,15.25) .. controls (219.5,26.75) and (254,65.75) .. (210.5,53.75) .. controls (167,41.75) and (116,78.75) .. (96,48.75) .. controls (76,18.75) and (99.5,4.75) .. (123,3.25) -- cycle ;
\draw   (3.5,153.63) .. controls (3.5,134.23) and (29.69,118.5) .. (62,118.5) .. controls (94.31,118.5) and (120.5,134.23) .. (120.5,153.63) .. controls (120.5,173.03) and (94.31,188.76) .. (62,188.76) .. controls (29.69,188.76) and (3.5,173.03) .. (3.5,153.63) -- cycle ;
\draw  [draw opacity=0] (98.77,146.93) .. controls (96.97,158.82) and (81.37,168.11) .. (62.38,168.11) .. controls (44.42,168.11) and (29.49,159.79) .. (26.41,148.83) -- (62.38,144.63) -- cycle ; \draw   (98.77,146.93) .. controls (96.97,158.82) and (81.37,168.11) .. (62.38,168.11) .. controls (44.42,168.11) and (29.49,159.79) .. (26.41,148.83) ;
\draw  [draw opacity=0] (30.97,156.86) .. controls (35.88,148.3) and (47.67,142.28) .. (61.42,142.28) .. controls (76.12,142.28) and (88.57,149.16) .. (92.79,158.65) -- (61.42,165.76) -- cycle ; \draw   (30.97,156.86) .. controls (35.88,148.3) and (47.67,142.28) .. (61.42,142.28) .. controls (76.12,142.28) and (88.57,149.16) .. (92.79,158.65) ;
\draw   (244,153.08) .. controls (244,120.04) and (262.69,93.25) .. (285.75,93.25) .. controls (308.81,93.25) and (327.5,120.04) .. (327.5,153.08) .. controls (327.5,186.13) and (308.81,212.92) .. (285.75,212.92) .. controls (262.69,212.92) and (244,186.13) .. (244,153.08) -- cycle ;
\draw  [draw opacity=0] (303.13,151.74) .. controls (299.89,163.97) and (293.66,172.25) .. (286.5,172.25) .. controls (279.21,172.25) and (272.87,163.65) .. (269.69,151.04) -- (286.5,132.5) -- cycle ; \draw   (303.13,151.74) .. controls (299.89,163.97) and (293.66,172.25) .. (286.5,172.25) .. controls (279.21,172.25) and (272.87,163.65) .. (269.69,151.04) ;
\draw  [draw opacity=0] (273.16,161.82) .. controls (275.93,153.53) and (280.61,148.08) .. (285.91,148.08) .. controls (291.18,148.08) and (295.83,153.46) .. (298.61,161.66) -- (285.91,179.42) -- cycle ; \draw   (273.16,161.82) .. controls (275.93,153.53) and (280.61,148.08) .. (285.91,148.08) .. controls (291.18,148.08) and (295.83,153.46) .. (298.61,161.66) ;
\draw    (132,150.01) -- (223.5,150.24) ;
\draw [shift={(225.5,150.25)}, rotate = 180.15] [color={rgb, 255:red, 0; green, 0; blue, 0 }  ][line width=0.75]    (10.93,-3.29) .. controls (6.95,-1.4) and (3.31,-0.3) .. (0,0) .. controls (3.31,0.3) and (6.95,1.4) .. (10.93,3.29)   ;
\draw [shift={(130,150)}, rotate = 0.15] [color={rgb, 255:red, 0; green, 0; blue, 0 }  ][line width=0.75]    (10.93,-3.29) .. controls (6.95,-1.4) and (3.31,-0.3) .. (0,0) .. controls (3.31,0.3) and (6.95,1.4) .. (10.93,3.29)   ;
\draw    (110,64.75) -- (70.79,111.22) ;
\draw [shift={(69.5,112.75)}, rotate = 310.15999999999997] [color={rgb, 255:red, 0; green, 0; blue, 0 }  ][line width=0.75]    (10.93,-3.29) .. controls (6.95,-1.4) and (3.31,-0.3) .. (0,0) .. controls (3.31,0.3) and (6.95,1.4) .. (10.93,3.29)   ;
\draw    (211.5,63.75) -- (244.37,111.6) ;
\draw [shift={(245.5,113.25)}, rotate = 235.52] [color={rgb, 255:red, 0; green, 0; blue, 0 }  ][line width=0.75]    (10.93,-3.29) .. controls (6.95,-1.4) and (3.31,-0.3) .. (0,0) .. controls (3.31,0.3) and (6.95,1.4) .. (10.93,3.29)   ;
\draw (150.5,155) node [anchor=north west][inner sep=0.75pt]  [font=\small] [align=left] {T-duality};
\draw (26.5,193) node [anchor=north west][inner sep=0.75pt]  [font=\footnotesize] [align=left] {bundle gerbe};
\draw (249.5,216.5) node [anchor=north west][inner sep=0.75pt]  [font=\footnotesize] [align=left] {bundle gerbe};
\draw (61.5,70.5) node [anchor=north west][inner sep=0.75pt]   [align=left] {{\footnotesize atlas}};
\draw (229,70.5) node [anchor=north west][inner sep=0.75pt]   [align=left] {{\footnotesize atlas}};
\draw (111.5,-13) node [anchor=north west][inner sep=0.75pt]  [font=\footnotesize] [align=left] {"doubled space"};
\end{tikzpicture}
\caption{the "doubled space" seen as the atlas of both a bundle gerbe and its T-dual.}
\end{center}\end{figure}

\newpage

\noindent If two different bundle gerbes $\mathscr{G}$ and $\widetilde{\mathscr{G}}$ are T-dual, they will have the same atlas $\M$. In other words, we will have a correspondence
\begin{equation}
    \begin{tikzcd}[row sep=3ex, column sep=3ex] 
    & \M \arrow[dr, "\widetilde{\Phi}"]\arrow[dl, "\Phi"']& \\
    \mathscr{G} & & \widetilde{\mathscr{G}}.
\end{tikzcd}
\end{equation}
This can be seen directly by looking at the \v{C}ech data of the T-dual bundle gerbes with the respective connections, as it was done in \cite{Alf19, Alf20}. It is not hard to see it from the isomorphism $\mathscr{G}\times_{M_0}\widetilde{M} \cong M\times_{M_0}\widetilde{\mathscr{G}}$ underlying T-duality between $\mathscr{G}\twoheadrightarrow M$ and $\widetilde{\mathscr{G}}\twoheadrightarrow\widetilde{M}$.
\vspace{0.25cm}

\noindent As we will show in the next subsection, the lift of the T-duality to the atlas will be an isometry $(\mathcal{M},J,\omega)\longrightarrow(\mathcal{M},\widetilde{J},\widetilde{\omega})$ of para-Hermitian manifolds, i.e. a smooth map which preserves the para-Hermitian metric $\eta(-,-):=\omega(J-,-)$ of $\mathcal{M}$.

\subsection{Topological T-duality}

\begin{theorem}[Topological T-duality on the doubled space]
Let $\mathscr{G}\xtwoheadrightarrow{\;\Pi\;} M$ and $\widetilde{\mathscr{G}}\xtwoheadrightarrow{\;\widetilde{\Pi}\;} \widetilde{M}$ be two bundle gerbes equipped with connection, such that they are T-dual. Then their atlases, respectively $(\M,J,\omega)$ and $(\M,\widetilde{J},\widetilde{\omega})$, are related by a para-Hermitian isometry, i.e. a change of polarisation as defined by \cite{MarSza18}.
\end{theorem}

\newpage

\begin{proof}
Let us start from the T-duality diagram of two topologically T-dual bundle gerbes. An atlas will sit on top of the diagram as it follows:
\begin{equation}
    \begin{tikzcd}[row sep={11ex,between origins}, column sep={12ex,between origins}]
    & & \M \arrow[dr, shorten <= 0.1em, shorten >= 0.1em]\arrow[ld, shorten <= 0.1em, shorten >= 0.1em]\arrow[ddll, bend right=50, "\Phi"']\arrow[ddrr, bend left=50, "\widetilde{\Phi}"] & & \\
    & \mathscr{G}\times_{M_0}\widetilde{M} \arrow[dr, "\Pi"', shorten <= 0.1em, shorten >= 0.1em]\arrow[dl, "\pi", shorten <= 0.1em, shorten >= 0.1em] & & M\times_{M_0}\widetilde{\mathscr{G}} \arrow[dr, "\widetilde{\pi}"', shorten <= 0.1em, shorten >= 0.1em]\arrow[dl, "\widetilde{\Pi}", shorten <= 0.1em, shorten >= 0.1em] \\
    \mathscr{G} \arrow[dr, "\Pi"', shorten <= 0.1em, shorten >= 0.1em] & & M\times_{M_0}\widetilde{M} \arrow[dr, "\pi"', shorten <= 0.1em, shorten >= 0.1em]\arrow[dl, "\widetilde{\pi}", shorten <= 0.1em, shorten >= 0.1em] & & [-2.5em]\widetilde{\mathscr{G}} \arrow[dl, "\widetilde{\Pi}", shorten <= 0.1em, shorten >= 0.1em] \\
    & M \arrow[dr, "\pi"', shorten <= 0.1em, shorten >= 0.1em] & & \widetilde{M} \arrow[dl, "\widetilde{\pi}", shorten <= 0.1em, shorten >= 0.1em] & \\
    & & M_0 & &
    \end{tikzcd}
\end{equation}
Let us consider the atlas $(\M,J,\omega)$ of the bundle gerbe $\mathscr{G}\twoheadrightarrow M$. Let $e^i\in\Omega^1(M)$ be the connection of the $T^n$-bundle $M\twoheadrightarrow M_0$. As shown in \cite[p.$\,$46]{Alf19}, we can expand the local $2$-form potential of the bundle gerbe in the connection $e^i\in\Omega^1(M)$ by 
\begin{equation}
    B_{(\alpha)} \;=\; B^{(0)}_{ij}e^i \wedge e^j + B^{(1)}_{(\alpha)\mu i}\di x^\mu \wedge e^i + {B}^{(2)}_{(\alpha)\mu\nu}\di x^\mu \wedge \di x^\nu
\end{equation}
where $B^{(0)}_{ij}$ is a globally defined scalar moduli field on $M$ and, therefore, we omitted the $\alpha$-index.
The corresponding fundamental $2$-form on the atlas $\M$ will be
\begin{equation}
    \omega \;=\; \big(\widetilde{e}_i+ B^{(0)}_{ij}e^j\big) \wedge e^i  + \widetilde{e}_\mu\wedge e^\mu,
\end{equation}
where we patch-wise defined the following global $1$-forms on the atlas:
\begin{equation}
    \begin{aligned}
        e^\mu \;&=\; \di x^\mu &\quad e^i \;&=\; \di\theta_{(\alpha)}^i + A^i_{(\alpha)\mu}\di x^\mu \\[0.4em]
        \widetilde{e}_{\mu} \;&=\; \di \widetilde{x}_{(\alpha)\mu}+ B^{(2)}_{(\alpha)\mu\nu}\di x^\nu \quad & \widetilde{e}_{i} \;&=\; \di\widetilde{\theta}_{(\alpha)i} + B^{(1)}_{(\alpha)i\mu}\di x^\mu
    \end{aligned}
\end{equation}
Let us explicitly construct the para-Hermitian metric $\eta$ of the atlas. This will globally be
\begin{equation}
    \eta(-,-) \;:=\; \omega(J-,-) \quad\; \Rightarrow \quad\; \eta \;=\; \widetilde{e}_i \odot e^i + \widetilde{e}_\mu \odot e^\mu
\end{equation}
Since $b:=B^{(0)}_{ij}e^i\wedge e^j\in\Omega^2(M)$ is a global $2$-form, the moduli field $B^{(0)}_{ij}\in\Coo(M,\mathfrak{so}(n))$ can be interpreted as a global B-shift. Thus, there exists an isometry of our para-Hermitian manifold \cite[p.$\,$15]{MarSza18} given by
\begin{equation}
    \omega' \;=\; e^b\,\omega \;=\; \widetilde{e}_i \wedge e^i  + \widetilde{e}_\mu\wedge e^\mu,
\end{equation}
By using this isometry, we forgot the moduli field and we retained only the topologically relevant component of the connection.
Now, let $(\M,\widetilde{J},\widetilde{\omega})$ be the atlas of the bundle gerbe $\widetilde{\mathscr{G}}\twoheadrightarrow \widetilde{M}$. Since we started from a couple of T-dual geometric backgrounds ${\mathscr{G}}$ and $\widetilde{\mathscr{G}}$, we already know that the potential $2$-form of the latter is
\begin{equation}
    \widetilde{B}_{(\alpha)} \;=\; \widetilde{B}^{(0)ij}\widetilde{e}_i \wedge \widetilde{e}_j + A_{(\alpha)\mu}^i\di x^\mu \wedge \widetilde{e}_i + {B}^{(2)}_{(\alpha)\mu\nu}\di x^\mu \wedge \di x^\nu
\end{equation}
where $\widetilde{B}^{(0)ij}$ is a global moduli field (which can be explicitly obtained by using the Buscher rules) and $A_{(\alpha)\mu}^i$ is the $1$-form potential of the $T^n$-bundle $M\twoheadrightarrow M_0$. Therefore, the T-dual corresponding fundamental $2$-form will be
\begin{equation}
    \widetilde{\omega} \;=\; \big(e^i+\widetilde{B}^{(0)ij}\widetilde{e}_j\big)\wedge \widetilde{e}_i + \widetilde{e}_\mu\wedge e^\mu
\end{equation}
Similarly to the first bundle gerbe, $\widetilde{b}:=\widetilde{B}^{(0)ij}\widetilde{e}_i\wedge\widetilde{e}_j$ is a global $2$-form and, thus, the map
\begin{equation}
    \widetilde{\omega}' \;=\; e^{\widetilde{b}}\,\widetilde{\omega} \;=\; e^i\wedge \widetilde{e}_i + \widetilde{e}_\mu\wedge e^\mu 
\end{equation}
is an isometry of the para-Hermitian metric. 
Now, let us call $J'$ and $\widetilde{J}'$ the para-complex structures corresponding to $\omega'$ and $\widetilde{\omega}'$. We need to find a morphism of para-Hermitian manifolds $f:(\M,J',\omega')\longrightarrow(\M,\widetilde{J}',\widetilde{\omega}')$ such that $\widetilde{\omega}' \;=\; f^\ast\omega'$ and check that it is an isometry. This is immediately the map  $f:\big(x_{(\alpha)},\widetilde{x}_{(\alpha)},\theta_{(\alpha)},\widetilde{\theta}_{(\alpha)}\big)\mapsto \big(x_{(\alpha)},\widetilde{x}_{(\alpha)},\widetilde{\theta}_{(\alpha)},\theta_{(\alpha)}\big)$, which is given by the exchange of the torus coordinates $\theta$ and $\widetilde{\theta}$ on each chart and is clearly an isometry. Therefore, by composition, we obtained an isometry $e^{b}\circ f\circ e^{-\widetilde{b}}:(\M,J,\omega)\longrightarrow(\M,\widetilde{J},\widetilde{\omega})$.
\end{proof}

\begin{remark}[Buscher rules]
In \cite[p.$\,$47]{Alf19}, we also showed that the Buscher transformations $(g^{(0)}_{ij}, B^{(0)}_{ij})\mapsto(\widetilde{g}^{(0)ij}, \widetilde{B}^{(0)ij})$ of the moduli field of the metric and the Kalb-Ramond field follow directly from applying the isometry of the lemma to the generalised metric, i.e. $\widetilde{\mathcal{G}}=f^\ast\mathcal{G}$.
\end{remark}

\subsection{Non-geometric T-duality}

We identified the isometries of our atlas $(\M,J,\omega)$ with changes of polarisation, i.e. with changes of T-duality frame. However, in general, it is not be possible to identify the target $(\M,\widetilde{J},\widetilde{\omega})$ of an isometry with the atlas of another bundle gerbe. In general, we can also obtain an almost para-complex structure $\widetilde{J}$ which is not integrable. In this case, the background described by the transformed atlas $(\M,\widetilde{J},\widetilde{\omega})$ is, then, a non-geometric background. \vspace{0.25cm}

\noindent For a detailed discussion of the non-geometric cases, such as T-folds, in the context of higher geometry and atlases we redirect to \cite{Alf19, Alf20}.

\section{Outlook}

The correspondence between doubled spaces and bundle gerbes we explored in this paper sheds new light on the global geometry underlying Double Field Theory. Moreover, it provides a higher geometric explanation for the appearance of the extra coordinates and for para-Hermitian geometry. These results are particularly important for the investigation of the other extended geometries, i.e. the exceptional geometries underlying Exceptional Field Theories, whose globalisation is significantly more obscure. In particular, the higher geometric perspective will allow to find a generalisation of para-Hermitian geometry for Exceptional Field Theory. Even if exceptional generalised geometry \cite{Wald08E, Wald11, Wald12} is well-understood, such a generalisation is still completely unknown. A generalised para-Hermitian formalism would be extremely fruitful, for example, in the current research in exceptional Drinfel'd geometries\cite{Sakatani:2019zrs, Malek:2019xrf, Blair:2020ndg, Musaev:2020nrt, Sakatani:2020wah, Malek:2020hpo}. \vspace{0.15cm}

\noindent In \cite{Cederwall:2017fjm, Cederwall:2018aab, Cederwall:2018kqk}, extended geometry has been studied in algebraic terms, in the light of representation theory. The extended/higher correspondence will then provide a complementary global geometric perspective to extended geometry, as well as new connections between higher geometry and representation theory.
\vspace{0.15cm}

\noindent Moreover, since the non-perturbative quantisation of strings and branes can be achieved by higher geometric quantisation \cite{SaSza11,BSS16,BS16, FSS16}, the close relation we established between extended and higher geometries will have a profound impact on the problem of quantisation. This issue, among other ones, was started to be studied in \cite{Alfonsi:2021bot}.
\vspace{0.15cm}

\noindent The higher structure which encompasses the global geometry of the $C$-field of $11$-dimensional supergravity can be seen as a bundle $5$-gerbe twisted by a bundle $2$-gerbe \cite{FSS15x}, which gives rise to the following diagram:
\begin{equation}
\begin{tikzcd}[row sep=10ex, column sep=7ex]
\mathscr{G}_{\mathrm{M5}}\arrow[r]\arrow[d, "\Pi_{\mathrm{M5}}"] & \ast \arrow[d] & \\
\mathscr{G}_{\mathrm{M2}}\arrow[r, "f_{\mathrm{M5}}"]\arrow[d, "\Pi_{\mathrm{M2}}"] & \mathbf{B}^6U(1) \arrow[r]\arrow[d] & \ast \arrow[d]\\
M \arrow[r, "f_{\mathrm{M2/5}}"]\arrow[rr, "f_{\mathrm{M2}}", bend right=25]& \mathbf{B}^6U(1)/\!/\mathbf{B}^2U(1), \arrow[r] & \mathbf{B}^3U(1),
\end{tikzcd}
\end{equation}
where the twisted cocycle $f_{\mathrm{M2/5}}$ can be also generalised to a $4$-cohomotopy cocycle $M\xrightarrow{f_{\mathrm{M2/5}}}S^4$, where the $4$-sphere can be given in terms of its minimal Sullivan dg-algebra by
\begin{equation}\label{sphere}
    \mathrm{CE}(S^4)\;=\; \mathbb{R}[g_4,g_7]/\langle\di g_4=0,\;\di g_7 + g_4\wedge g_4=0\rangle,
\end{equation}
where $g_4$ and $g_7$ are respectively $4$- and $7$-degree generators.
In the context of $L_\infty$-superalgebras, a notion of super exceptional space $\mathbb{R}^{1,10|\mathbf{32}}_{\mathrm{ex}}$ has been defined by \cite{FSS18, FSS19x, FSS20}. Notice that, in its bosonic form, i.e.
\begin{equation}
    \mathbb{R}^{1,10}_{\mathrm{ex}} \;=\;  \mathbb{R}^{1,10}\oplus \wedge^2(\mathbb{R}^{1,10})^\ast \oplus\wedge^5(\mathbb{R}^{1,10})^\ast  ,
\end{equation}
can also be interpreted as the atlas of the linearised version of the twisted bundle $5$-gerbe $\mathscr{G}_\mathrm{M5}\twoheadrightarrow M$.
If we split the base space in time and space by $\mathbb{R}^{1,10}=\mathbb{R}^{1}_{\mathrm{t}}\oplus \mathbb{R}^{10}$, we obtain the decomposition
\begin{equation}
    \mathbb{R}^{1,10}_{\mathrm{ex}} \;=\; \underbrace{ \mathbb{R}^{10}}_{\text{pp-wave}}\!\oplus\, \underbrace{\wedge^2(\mathbb{R}^{10})^\ast}_{\text{M2-brane}} \oplus \underbrace{ \wedge^2\mathbb{R}^{10} }_{\text{M9-brane}} \oplus \underbrace{\wedge^5(\mathbb{R}^{10})^\ast}_{\text{M5-brane}} \,\oplus\!\!\!  \underbrace{\wedge^6\mathbb{R}^{10}}_{\text{KK-monopole}} ,
\end{equation}
which agrees with the description of brane charges in M-theory \cite{Hull:1997kt}.
Notice that, if we split the base space in an internal and external space by $\mathbb{R}^{1,10}=\mathbb{R}^{1,3}\oplus \mathbb{R}^{7}$, we obtain
\begin{equation}
    \mathbb{R}^{1,10}_{\mathrm{ex}} \;=\; \mathbb{R}^{1,3} \oplus \Big(\mathbb{R}^{7}\oplus \wedge^2(\mathbb{R}^{7})^\ast \oplus\wedge^5(\mathbb{R}^{7})^\ast \oplus \wedge^6\mathbb{R}^{7} \Big) \oplus \,\cdots ,
\end{equation}
where the terms we explicitly wrote correspond to the $(4+56)$-dimensional extended space underlying $E_{7(7)}$ Exceptional Field Theory \cite{Hohm:2013uia}. Moreover, the terms we omitted are mixed terms involving wedge products between $\mathbb{R}^{1,3}$ and $\mathbb{R}^{7}$ which correspond to tensor hierarchies \cite{Cagnacci:2018buk, Hohm19DFT} at $0$-degree. 
Moreover, as already argued by \cite[sec.$\,$9.2]{Arvanitakis:2018hfn}, the naturally expected structure generalising the fundamental $2$-form to the exceptional case would generally be an almost $n$-plectic structure. Recently, \cite{Sakatani:2020umt} proposed a local generalisation of the Born $\sigma$-model of the string to the M-branes. These are equipped with $3$- and $6$-forms which appear to be closely related to the transgression of the higher field whose curvature comes from the dg-algebra \eqref{sphere}.
All these are strong hints that the correspondence between extended geometry and higher geometry via atlases can be well-defined for the exceptional cases too.


\section*{Acknowledgement}
I want to thank the organisers Vicente Cortés, Liana David and Carlos Shahbazi of the workshop \href{https://www.math.uni-hamburg.de/projekte/gg2020/}{\textit{Generalised Geometry and Applications 2020}} at Universit\"{a}t Hamburg. I would like to thank Christian S\"{a}mann, Franco Pezzella, Emanuel Malek, Urs Schreiber, Richard Szabo and Francesco Genovese for fruitful discussion. I want also to thank Chris Blair and Yuho Sakatani for interesting comments on the preprint of this paper. Finally, I would like to thank the reviewer, who provided .

\medskip
\addcontentsline{toc}{section}{References}
\bibliographystyle{fredrickson}
{\footnotesize \singlespacing
\bibliography{sample}
}
\end{document}